%% file: DESY-13-054.tex
\documentclass[zpreprint,zbstdefault,british]{zeus_paper}
\usepackage{zeusdet_units}
\input zeus_defs.tex
\input DESY-13-054-cit.tex
\begin{document}
\input{DESY-13-054-tit}
\thispagestyle{empty}
\clearpage
\input{DESY-13-054-aut}
\clearpage
\pagenumbering{arabic}
%
\include{DESY-13-054-txt}

\include{DESY-13-054-ref}
\include{DESY-13-054-tab}
\include{DESY-13-054-fig}

%
%
\end{document}

%% file: zeus_defs.tex
\chardef\usc=95
\chardef\til=126
\catcode`\@=11 
\DeclareRobustCommand\xdotspace{\futurelet\@let@token\@xdotspace}
\def\@xdotspace{%
  \ifx\@let@token.\else
  \ifx\@let@token\bgroup.\else
  \ifx\@let@token\egroup.\else
  \ifx\@let@token\/.\else
  \ifx\@let@token\ .\else
  \ifx\@let@token~.\else
  \ifx\@let@token!.\else
  \ifx\@let@token,.\else
  \ifx\@let@token:.\else
  \ifx\@let@token;.\else
  \ifx\@let@token?.\else
  \ifx\@let@token/.\else
  \ifx\@let@token'.\else
  \ifx\@let@token).\else
  \ifx\@let@token-.\else
  \ifx\@let@token\@xobeysp.\else
  \ifx\@let@token\space.\else
  \ifx\@let@token\@sptoken.\else
   .\space
   \fi\fi\fi\fi\fi\fi\fi\fi\fi\fi\fi\fi\fi\fi\fi\fi\fi\fi}
\catcode`\@=12 

\newcommand{\stru}[2]{%
   \relax\ifmmode\hbox{\vrule height#1 depth#2 width0pt}%
   \else\vrule height#1 depth#2 width0pt\fi}

\newcommand{\Ronum}[1]{\uppercase\expandafter{\romannumeral#1}}
\newcommand{\ronum}[1]{\expandafter{\romannumeral#1}}
\DeclareRobustCommand{\LaTeXZ}{%
  \LaTeX\kern-.05em4\kern-.1em
  {\raisebox{-0.2ex}{$\scriptstyle\text{ZEUS}$}}\xspace}

\newcommand{\slashfrac}[2]{%
  \raisebox{0.5ex}{\ensuremath #1}\kern-0.12em/\kern-0.08em
  \raisebox{-.8ex}{\ensuremath #2}}

\newcommand{\sqr}[3]{%
    {\vcenter{\hrule height.#3ex\hbox{\vrule width.#2ex height#1ex
     \kern#1ex\vrule width.#3ex}\hrule height.#2ex}}}

\catcode`\@=11 
\newcommand{\parenbar}{\mathpalette\p@renb@r}
\def\p@renb@r#1#2{\vbox{%
  \ifx#1\scriptscriptstyle \dimen@.7em\dimen@ii.2em\else
  \ifx#1\scriptstyle \dimen@.8em\dimen@ii.25em\else
  \dimen@1em\dimen@ii.4em\fi\fi \offinterlineskip
  \ialign{\hfill##\hfill\cr
    \vbox{\hrule width\dimen@ii}\cr
    \noalign{\vskip-.3ex}%
    \hbox to\dimen@{$\mathchar300\hfil\mathchar301$}\cr
    \noalign{\vskip-.3ex}%
    $#1#2$\cr}}}
\catcode`\@=12 

\ifzeusunit
  
\else
  
\fi

\newcommand{\IP}{{\rm I$\kern-0.01667em$P}\xspace}

\mathchardef\qsm=63
\mathchardef\pls=43
\mathchardef\mns=512
\mathchardef\plm=518
\mathchardef\eql=61
\mathchardef\smallleft=300
\mathchardef\smallright=301
\mathchardef\les=316
\mathchardef\gre=318
\mathchardef\leq=532
\mathchardef\grq=533

\catcode`\@=11 
\newcounter{pict@width}
\newcounter{pict@height}
\newlength{\pict@scale}
\newcommand{\psfigadd}[4]{%
\setcounter{pict@width}{1*\ratio{#2+\pict@scale/2}{\pict@scale}}
\setcounter{pict@height}{1*\ratio{#3+\pict@scale/2}{\pict@scale}}
\setlength{\unitlength}{\pict@scale}
\hbox to #2{\hspace{-\fill}\begin{picture}(\thepict@width,\thepict@height)
\put(0,0){\psfig{figure=#1,width=#2,height=#3,clip=}}
\SetScale{0.283466457}
\SetWidth{1.763889}
{#4}
\end{picture}}
}
\newcounter{pict@widthfst}
\newcounter{pict@widthscd}
\newcounter{pict@widthtot}
\newcommand{\psfigaddtwo}[7]{%
\setcounter{pict@widthfst}{1*\ratio{#2+\pict@scale/2}{\pict@scale}}
\setcounter{pict@widthscd}{1*\ratio{#2+#4+\pict@scale/2}{\pict@scale}}
\setcounter{pict@widthtot}{1*\ratio{#2+#4+#6+\pict@scale/2}{\pict@scale}}
\setcounter{pict@height}{1*\ratio{#3+\pict@scale/2}{\pict@scale}}
\setlength{\unitlength}{\pict@scale}
\hbox{\hspace{-\fill}\begin{picture}(\thepict@widthtot,\thepict@height)
\put(0,0){\psfig{figure=#1,width=#2,height=#3,clip=}}
\put(\thepict@widthscd,0){\psfig{figure=#5,width=#6,height=#3,clip=}}
\SetScale{0.283466457}
\SetWidth{1.763889}
{#7}
\end{picture}}
}
\newcommand{\psfigror}[4]{%
\setcounter{pict@width}{1*\ratio{#2+\pict@scale/2}{\pict@scale}}
\setcounter{pict@height}{1*\ratio{#3+\pict@scale/2}{\pict@scale}}
\setlength{\unitlength}{\pict@scale}
\hbox{\begin{picture}(\thepict@width,\thepict@height)
\put(0,\thepict@height){\psfig{figure=#1,width=#3,height=#2,clip=,angle=270}}
\SetScale{0.283466457}
\SetWidth{1.763889}
{#4}
\end{picture}}
}
\newcommand{\psfigrol}[4]{%
\setcounter{pict@width}{1*\ratio{#2+\pict@scale/2}{\pict@scale}}
\setcounter{pict@height}{1*\ratio{#3+\pict@scale/2}{\pict@scale}}
\setlength{\unitlength}{\pict@scale}
\hbox{\begin{picture}(\thepict@width,\thepict@height)
\put(0,0){\psfig{figure=#1,width=#3,height=#2,clip=,angle=90}}
\SetScale{0.283466457}
\SetWidth{1.763889}
{#4}
\end{picture}}
}
\catcode`\@=12 
\newlength\listtextwidth

\catcode`\@=11 
\newlength{\@tabfninsert}
\newlength{\@tabfnwidth}
\newcommand{\tabfootnote}[2]{%
  \setlength{\@tabfninsert}{0.8em}
  \setlength{\@tabfnwidth}{\textwidth}
  \addtolength{\@tabfnwidth}{-\@tabfninsert}
  \addtolength{\@tabfnwidth}{-0.4em}
  \noindent\makebox[\@tabfninsert][r]{\footnotesize$^{#1}$\hfil}\hfill%
  \parbox[t]{\@tabfnwidth}{\footnotesize #2\hfill}}
\catcode`\@=12 

%% file: DESY-13-054-cit.tex
\def\citeCTD{{\cite{%
nim:a279:290,*npps:b32:181,*nim:a338:254%
}}\xspace}
\def\citeMVD{{\cite{%
nim:a581:656%
}}\xspace}
\def\citeSTT{{\cite{%
nim:a535:191%
}}\xspace}
\def\citeCAL{{\cite{%
nim:a309:77,*nim:a309:101,*nim:a321:356,*nim:a336:23%
}}\xspace}
\def\citeHPDF{
{\cite{%
Herapdf1.5:2010,*Radescu:2010zz%
}}\xspace}
\def\citeDsPhP{
{\cite{%
pl:b401:192,*epj:c6:67%
}}\xspace}
\def\citeHES{{\cite{%
nim:a277:176%
}}\xspace}
\def\citeSixmT{{\cite{%
thesis:gosau:2007%
}}\xspace}
\def\citeBAC{{\cite{%
nim:a300:480%
}}\xspace}
\def\citePCAL{{\cite{%
desy-92-066,*zfp:c63:391,*acpp:b32:2025%
}}\xspace}
\def\citeSPECTRO{{\cite{%
nim:a565:572%
}}\xspace}

%% file: DESY-13-054-tit.tex
\prepnum{DESY--13--054}

\title{
  Measurement of $ \mathbf{D^{*\pm}}$ production in deep inelastic scattering at HERA
}                   
\author{ZEUS Collaboration}
\abstract{
The production of $D^{*\pm}$ mesons in deep inelastic $ep$ scattering has been measured
for exchanged photon virtualities $ 5<Q^2<1000 \gev^2 $,
using an integrated luminosity of $363$~pb$^{-1}$ with the ZEUS detector at HERA.
Differential cross sections have been measured and compared to
next-to-leading-order QCD calculations.
The cross-sections are used to extract the charm contribution to
the proton structure functions, expressed in terms of the reduced
charm cross section, $\sigma_{\rm red}^{c\bar{c}}$.
Theoretical calculations based on fits to inclusive HERA data are compared to the results.
}
\makezeustitle

%% file: DESY-13-054-aut.tex
%
%
%
%

\pagenumbering{Roman}
                                                   %
\begin{center}
{                      \Large  The ZEUS Collaboration              }
\end{center}

{\small


        {\raggedright
H.~Abramowicz$^{45, aj}$, 
I.~Abt$^{35}$, 
L.~Adamczyk$^{13}$, 
M.~Adamus$^{54}$, 
R.~Aggarwal$^{7, c}$, 
S.~Antonelli$^{4}$, 
P.~Antonioli$^{3}$, 
A.~Antonov$^{33}$, 
M.~Arneodo$^{50}$, 
O.~Arslan$^{5}$
% ...
        }

\newpage


\makebox[3em]{$^{1}$}
\begin{minipage}[t]{14cm}
{\it Argonne National Laboratory, Argonne, Illinois 60439-4815, USA}~$^{A}$

\end{minipage}\\
\makebox[3em]{$^{2}$}
\begin{minipage}[t]{14cm}
{\it Andrews University, Berrien Springs, Michigan 49104-0380, USA}

\end{minipage}\\
\makebox[3em]{$^{3}$}
\begin{minipage}[t]{14cm}
{\it INFN Bologna, Bologna, Italy}~$^{B}$

\end{minipage}\\
\makebox[3em]{$^{4}$}
\begin{minipage}[t]{14cm}
{\it University and INFN Bologna, Bologna, Italy}~$^{B}$

\end{minipage}\\
\makebox[3em]{$^{5}$}
\begin{minipage}[t]{14cm}
{\it Physikalisches Institut der Universit\"at Bonn,
Bonn, Germany}~$^{C}$

\end{minipage}\\
\makebox[3em]{$^{6}$}
\begin{minipage}[t]{14cm}
{\it H.H.~Wills Physics Laboratory, University of Bristol,
Bristol, United Kingdom}~$^{D}$

\end{minipage}\\
\makebox[3em]{$^{7}$}
\begin{minipage}[t]{14cm}
{\it Panjab University, Department of Physics, Chandigarh, India}

\end{minipage}\\
\makebox[3em]{$^{8}$}
\begin{minipage}[t]{14cm}
{\it Calabria University,
Physics Department and INFN, Cosenza, Italy}~$^{B}$

\end{minipage}\\
\makebox[3em]{$^{9}$}
\begin{minipage}[t]{14cm}
{\it Institute for Universe and Elementary Particles, Chonnam National University,\\
Kwangju, South Korea}

\end{minipage}\\
\makebox[3em]{$^{10}$}
\begin{minipage}[t]{14cm}
{\it Jabatan Fizik, Universiti Malaya, 50603 Kuala Lumpur, Malaysia}~$^{E}$

\end{minipage}\\
\makebox[3em]{$^{11}$}
\begin{minipage}[t]{14cm}
{\it Nevis Laboratories, Columbia University, Irvington on Hudson,
New York 10027, USA}~$^{F}$

\end{minipage}\\
\makebox[3em]{$^{12}$}
\begin{minipage}[t]{14cm}
{\it The Henryk Niewodniczanski Institute of Nuclear Physics, Polish Academy of \\
Sciences, Krakow, Poland}~$^{G}$

\end{minipage}\\
\makebox[3em]{$^{13}$}
\begin{minipage}[t]{14cm}
{\it AGH-University of Science and Technology, Faculty of Physics and Applied Computer
Science, Krakow, Poland}~$^{H}$

\end{minipage}\\
\makebox[3em]{$^{14}$}
\begin{minipage}[t]{14cm}
{\it Department of Physics, Jagellonian University, Cracow, Poland}

\end{minipage}\\
\makebox[3em]{$^{15}$}
\begin{minipage}[t]{14cm}
{\it Deutsches Elektronen-Synchrotron DESY, Hamburg, Germany}

\end{minipage}\\
\makebox[3em]{$^{16}$}
\begin{minipage}[t]{14cm}
{\it Deutsches Elektronen-Synchrotron DESY, Zeuthen, Germany}

\end{minipage}\\
\makebox[3em]{$^{17}$}
\begin{minipage}[t]{14cm}
{\it INFN Florence, Florence, Italy}~$^{B}$

\end{minipage}\\
\makebox[3em]{$^{18}$}
\begin{minipage}[t]{14cm}
{\it University and INFN Florence, Florence, Italy}~$^{B}$

\end{minipage}\\
\makebox[3em]{$^{19}$}
\begin{minipage}[t]{14cm}
{\it Fakult\"at f\"ur Physik der Universit\"at Freiburg i.Br.,
Freiburg i.Br., Germany}

\end{minipage}\\
\makebox[3em]{$^{20}$}
\begin{minipage}[t]{14cm}
{\it School of Physics and Astronomy, University of Glasgow,
Glasgow, United Kingdom}~$^{D}$

\end{minipage}\\
\makebox[3em]{$^{21}$}
\begin{minipage}[t]{14cm}
{\it Department of Engineering in Management and Finance, Univ. of
the Aegean, Chios, Greece}

\end{minipage}\\
\makebox[3em]{$^{22}$}
\begin{minipage}[t]{14cm}
{\it Hamburg University, Institute of Experimental Physics, Hamburg,
Germany}~$^{I}$

\end{minipage}\\
\makebox[3em]{$^{23}$}
\begin{minipage}[t]{14cm}
{\it Imperial College London, High Energy Nuclear Physics Group,
London, United Kingdom}~$^{D}$

\end{minipage}\\
\makebox[3em]{$^{24}$}
\begin{minipage}[t]{14cm}
{\it Institute of Particle and Nuclear Studies, KEK,
Tsukuba, Japan}~$^{J}$

\end{minipage}\\
\makebox[3em]{$^{25}$}
\begin{minipage}[t]{14cm}
{\it Institute of Physics and Technology of Ministry of Education and
Science of Kazakhstan, Almaty, Kazakhstan}

\end{minipage}\\
\makebox[3em]{$^{26}$}
\begin{minipage}[t]{14cm}
{\it Institute for Nuclear Research, National Academy of Sciences, Kyiv, Ukraine}

\end{minipage}\\
\makebox[3em]{$^{27}$}
\begin{minipage}[t]{14cm}
{\it Department of Nuclear Physics, National Taras Shevchenko University of Kyiv, Kyiv, Ukraine}

\end{minipage}\\
\makebox[3em]{$^{28}$}
\begin{minipage}[t]{14cm}
{\it Kyungpook National University, Center for High Energy Physics, Daegu,
South Korea}~$^{K}$

\end{minipage}\\
\makebox[3em]{$^{29}$}
\begin{minipage}[t]{14cm}
{\it Institut de Physique Nucl\'{e}aire, Universit\'{e} Catholique de Louvain, Louvain-la-Neuve,\\
Belgium}~$^{L}$

\end{minipage}\\
\makebox[3em]{$^{30}$}
\begin{minipage}[t]{14cm}
{\it Departamento de F\'{\i}sica Te\'orica, Universidad Aut\'onoma
de Madrid, Madrid, Spain}~$^{M}$

\end{minipage}\\
\makebox[3em]{$^{31}$}
\begin{minipage}[t]{14cm}
{\it Department of Physics, McGill University,
Montr\'eal, Qu\'ebec, Canada H3A 2T8}~$^{N}$

\end{minipage}\\
\makebox[3em]{$^{32}$}
\begin{minipage}[t]{14cm}
{\it Meiji Gakuin University, Faculty of General Education,
Yokohama, Japan}~$^{J}$

\end{minipage}\\
\makebox[3em]{$^{33}$}
\begin{minipage}[t]{14cm}
{\it Moscow Engineering Physics Institute, Moscow, Russia}~$^{O}$

\end{minipage}\\
\makebox[3em]{$^{34}$}
\begin{minipage}[t]{14cm}
{\it Lomonosov Moscow State University, Skobeltsyn Institute of Nuclear Physics,
Moscow, Russia}~$^{P}$

\end{minipage}\\
\makebox[3em]{$^{35}$}
\begin{minipage}[t]{14cm}
{\it Max-Planck-Institut f\"ur Physik, M\"unchen, Germany}

\end{minipage}\\
\makebox[3em]{$^{36}$}
\begin{minipage}[t]{14cm}
{\it NIKHEF and University of Amsterdam, Amsterdam, Netherlands}~$^{Q}$

\end{minipage}\\
\makebox[3em]{$^{37}$}
\begin{minipage}[t]{14cm}
{\it Physics Department, Ohio State University,
Columbus, Ohio 43210, USA}~$^{A}$

\end{minipage}\\
\makebox[3em]{$^{38}$}
\begin{minipage}[t]{14cm}
{\it Department of Physics, University of Oxford,
Oxford, United Kingdom}~$^{D}$

\end{minipage}\\
\makebox[3em]{$^{39}$}
\begin{minipage}[t]{14cm}
{\it INFN Padova, Padova, Italy}~$^{B}$

\end{minipage}\\
\makebox[3em]{$^{40}$}
\begin{minipage}[t]{14cm}
{\it Dipartimento di Fisica dell' Universit\`a and INFN,
Padova, Italy}~$^{B}$

\end{minipage}\\
\makebox[3em]{$^{41}$}
\begin{minipage}[t]{14cm}
{\it Department of Physics, Pennsylvania State University, University Park,\\
Pennsylvania 16802, USA}~$^{F}$

\end{minipage}\\
\makebox[3em]{$^{42}$}
\begin{minipage}[t]{14cm}
{\it Polytechnic University, Tokyo, Japan}~$^{J}$

\end{minipage}\\
\makebox[3em]{$^{43}$}
\begin{minipage}[t]{14cm}
{\it Dipartimento di Fisica, Universit\`a 'La Sapienza' and INFN,
Rome, Italy}~$^{B}$

\end{minipage}\\
\makebox[3em]{$^{44}$}
\begin{minipage}[t]{14cm}
{\it Rutherford Appleton Laboratory, Chilton, Didcot, Oxon,
United Kingdom}~$^{D}$

\end{minipage}\\
\makebox[3em]{$^{45}$}
\begin{minipage}[t]{14cm}
{\it Raymond and Beverly Sackler Faculty of Exact Sciences, School of Physics, \\
Tel Aviv University, Tel Aviv, Israel}~$^{R}$

\end{minipage}\\
\makebox[3em]{$^{46}$}
\begin{minipage}[t]{14cm}
{\it Department of Physics, Tokyo Institute of Technology,
Tokyo, Japan}~$^{J}$

\end{minipage}\\
\makebox[3em]{$^{47}$}
\begin{minipage}[t]{14cm}
{\it Department of Physics, University of Tokyo,
Tokyo, Japan}~$^{J}$

\end{minipage}\\
\makebox[3em]{$^{48}$}
\begin{minipage}[t]{14cm}
{\it Tokyo Metropolitan University, Department of Physics,
Tokyo, Japan}~$^{J}$

\end{minipage}\\
\makebox[3em]{$^{49}$}
\begin{minipage}[t]{14cm}
{\it Universit\`a di Torino and INFN, Torino, Italy}~$^{B}$

\end{minipage}\\
\makebox[3em]{$^{50}$}
\begin{minipage}[t]{14cm}
{\it Universit\`a del Piemonte Orientale, Novara, and INFN, Torino,
Italy}~$^{B}$

\end{minipage}\\
\makebox[3em]{$^{51}$}
\begin{minipage}[t]{14cm}
{\it Department of Physics, University of Toronto, Toronto, Ontario,
Canada M5S 1A7}~$^{N}$

\end{minipage}\\
\makebox[3em]{$^{52}$}
\begin{minipage}[t]{14cm}
{\it Physics and Astronomy Department, University College London,
London, United Kingdom}~$^{D}$

\end{minipage}\\
\makebox[3em]{$^{53}$}
\begin{minipage}[t]{14cm}
{\it Faculty of Physics, University of Warsaw, Warsaw, Poland}

\end{minipage}\\
\makebox[3em]{$^{54}$}
\begin{minipage}[t]{14cm}
{\it National Centre for Nuclear Research, Warsaw, Poland}

\end{minipage}\\
\makebox[3em]{$^{55}$}
\begin{minipage}[t]{14cm}
{\it Department of Particle Physics and Astrophysics, Weizmann
Institute, Rehovot, Israel}

\end{minipage}\\
\makebox[3em]{$^{56}$}
\begin{minipage}[t]{14cm}
{\it Department of Physics, University of Wisconsin, Madison,
Wisconsin 53706, USA}~$^{A}$

\end{minipage}\\
\makebox[3em]{$^{57}$}
\begin{minipage}[t]{14cm}
{\it Department of Physics, York University, Ontario, Canada M3J 1P3}~$^{N}$

\end{minipage}\\
\vspace{30em} \pagebreak[4]


\makebox[3ex]{$^{ A}$}
\begin{minipage}[t]{14cm}
 supported by the US Department of Energy\
\end{minipage}\\
\makebox[3ex]{$^{ B}$}
\begin{minipage}[t]{14cm}
 supported by the Italian National Institute for Nuclear Physics (INFN) \
\end{minipage}\\
\makebox[3ex]{$^{ C}$}
\begin{minipage}[t]{14cm}
 supported by the German Federal Ministry for Education and Research (BMBF), under
 contract No. 05 H09PDF\
\end{minipage}\\
\makebox[3ex]{$^{ D}$}
\begin{minipage}[t]{14cm}
 supported by the Science and Technology Facilities Council, UK\
\end{minipage}\\
\makebox[3ex]{$^{ E}$}
\begin{minipage}[t]{14cm}
 supported by HIR and UMRG grants from Universiti Malaya, and an ERGS grant from the
 Malaysian Ministry for Higher Education\
\end{minipage}\\
\makebox[3ex]{$^{ F}$}
\begin{minipage}[t]{14cm}
 supported by the US National Science Foundation. Any opinion, findings and conclusions or
 recommendations expressed in this material are those of the authors and do not necessarily
 reflect the views of the National Science Foundation.\
\end{minipage}\\
\makebox[3ex]{$^{ G}$}
\begin{minipage}[t]{14cm}
 supported by the Polish Ministry of Science and Higher Education as a scientific project No.
 DPN/N188/DESY/2009\
\end{minipage}\\
\makebox[3ex]{$^{ H}$}
\begin{minipage}[t]{14cm}
 supported by the Polish Ministry of Science and Higher Education and its grants
 for Scientific Research\
\end{minipage}\\
\makebox[3ex]{$^{ I}$}
\begin{minipage}[t]{14cm}
 supported by the German Federal Ministry for Education and Research (BMBF), under
 contract No. 05h09GUF, and the SFB 676 of the Deutsche Forschungsgemeinschaft (DFG) \
\end{minipage}\\
\makebox[3ex]{$^{ J}$}
\begin{minipage}[t]{14cm}
 supported by the Japanese Ministry of Education, Culture, Sports, Science and Technology
 (MEXT) and its grants for Scientific Research\
\end{minipage}\\
\makebox[3ex]{$^{ K}$}
\begin{minipage}[t]{14cm}
 supported by the Korean Ministry of Education and Korea Science and Engineering Foundation\
\end{minipage}\\
\makebox[3ex]{$^{ L}$}
\begin{minipage}[t]{14cm}
 supported by FNRS and its associated funds (IISN and FRIA) and by an Inter-University
 Attraction Poles Programme subsidised by the Belgian Federal Science Policy Office\
\end{minipage}\\
\makebox[3ex]{$^{ M}$}
\begin{minipage}[t]{14cm}
 supported by the Spanish Ministry of Education and Science through funds provided by CICYT\
\end{minipage}\\
\makebox[3ex]{$^{ N}$}
\begin{minipage}[t]{14cm}
 supported by the Natural Sciences and Engineering Research Council of Canada (NSERC) \
\end{minipage}\\
\makebox[3ex]{$^{ O}$}
\begin{minipage}[t]{14cm}
 partially supported by the German Federal Ministry for Education and Research (BMBF)\
\end{minipage}\\
\makebox[3ex]{$^{ P}$}
\begin{minipage}[t]{14cm}
 supported by RF Presidential grant N 3920.2012.2 for the Leading Scientific Schools and by
 the Russian Ministry of Education and Science through its grant for Scientific Research on
 High Energy Physics\
\end{minipage}\\
\makebox[3ex]{$^{ Q}$}
\begin{minipage}[t]{14cm}
 supported by the Netherlands Foundation for Research on Matter (FOM)\
\end{minipage}\\
\makebox[3ex]{$^{ R}$}
\begin{minipage}[t]{14cm}
 supported by the Israel Science Foundation\
\end{minipage}\\
\vspace{30em} \pagebreak[4]


\makebox[3ex]{$^{ a}$}
\begin{minipage}[t]{14cm}
now at University of Salerno, Italy\
\end{minipage}\\
\makebox[3ex]{$^{ b}$}
\begin{minipage}[t]{14cm}
now at Queen Mary University of London, United Kingdom\
\end{minipage}\\
\makebox[3ex]{$^{ c}$}
\begin{minipage}[t]{14cm}
also funded by Max Planck Institute for Physics, Munich, Germany\
\end{minipage}\\
\makebox[3ex]{$^{ d}$}
\begin{minipage}[t]{14cm}
also Senior Alexander von Humboldt Research Fellow at Hamburg University,
 Institute of Experimental Physics, Hamburg, Germany\
\end{minipage}\\
\makebox[3ex]{$^{ e}$}
\begin{minipage}[t]{14cm}
also at Cracow University of Technology, Faculty of Physics,
 Mathematics and Applied Computer Science, Poland\
\end{minipage}\\
\makebox[3ex]{$^{ f}$}
\begin{minipage}[t]{14cm}
supported by the research grant No. 1 P03B 04529 (2005-2008)\
\end{minipage}\\
\makebox[3ex]{$^{ g}$}
\begin{minipage}[t]{14cm}
partially supported by the Polish National Science Centre projects DEC-2011/01/B/ST2/03643
 and DEC-2011/03/B/ST2/00220\
\end{minipage}\\
\makebox[3ex]{$^{ h}$}
\begin{minipage}[t]{14cm}
now at Rockefeller University, New York, NY
 10065, USA\
\end{minipage}\\
\makebox[3ex]{$^{ i}$}
\begin{minipage}[t]{14cm}
now at DESY group FS-CFEL-1\
\end{minipage}\\
\makebox[3ex]{$^{ j}$}
\begin{minipage}[t]{14cm}
now at Institute of High Energy Physics, Beijing, China\
\end{minipage}\\
\makebox[3ex]{$^{ k}$}
\begin{minipage}[t]{14cm}
now at DESY group FEB, Hamburg, Germany\
\end{minipage}\\
\makebox[3ex]{$^{ l}$}
\begin{minipage}[t]{14cm}
also at Moscow State University, Russia\
\end{minipage}\\
\makebox[3ex]{$^{ m}$}
\begin{minipage}[t]{14cm}
now at University of Liverpool, United Kingdom\
\end{minipage}\\
\makebox[3ex]{$^{ n}$}
\begin{minipage}[t]{14cm}
now at CERN, Geneva, Switzerland\
\end{minipage}\\
\makebox[3ex]{$^{ o}$}
\begin{minipage}[t]{14cm}
also affiliated with University College London, UK\
\end{minipage}\\
\makebox[3ex]{$^{ p}$}
\begin{minipage}[t]{14cm}
now at Goldman Sachs, London, UK\
\end{minipage}\\
\makebox[3ex]{$^{ q}$}
\begin{minipage}[t]{14cm}
also at Institute of Theoretical and Experimental Physics, Moscow, Russia\
\end{minipage}\\
\makebox[3ex]{$^{ r}$}
\begin{minipage}[t]{14cm}
also at FPACS, AGH-UST, Cracow, Poland\
\end{minipage}\\
\makebox[3ex]{$^{ s}$}
\begin{minipage}[t]{14cm}
partially supported by Warsaw University, Poland\
\end{minipage}\\
\makebox[3ex]{$^{ t}$}
\begin{minipage}[t]{14cm}
supported by the Alexander von Humboldt Foundation\
\end{minipage}\\
\makebox[3ex]{$^{ u}$}
\begin{minipage}[t]{14cm}
now at Istituto Nazionale di Fisica Nucleare (INFN), Pisa, Italy\
\end{minipage}\\
\makebox[3ex]{$^{ v}$}
\begin{minipage}[t]{14cm}
now at Haase Energie Technik AG, Neum\"unster, Germany\
\end{minipage}\\
\makebox[3ex]{$^{ w}$}
\begin{minipage}[t]{14cm}
now at Department of Physics, University of Bonn, Germany\
\end{minipage}\\
\makebox[3ex]{$^{ x}$}
\begin{minipage}[t]{14cm}
also affiliated with DESY, Germany\
\end{minipage}\\
\makebox[3ex]{$^{ y}$}
\begin{minipage}[t]{14cm}
also at University of Tokyo, Japan\
\end{minipage}\\
\makebox[3ex]{$^{ z}$}
\begin{minipage}[t]{14cm}
now at Kobe University, Japan\
\end{minipage}\\
\makebox[3ex]{$^{\dagger}$}
\begin{minipage}[t]{14cm}
 deceased \
\end{minipage}\\
\makebox[3ex]{$^{aa}$}
\begin{minipage}[t]{14cm}
supported by DESY, Germany\
\end{minipage}\\
\makebox[3ex]{$^{ab}$}
\begin{minipage}[t]{14cm}
member of National Technical University of Ukraine, Kyiv Polytechnic Institute,
 Kyiv, Ukraine\
\end{minipage}\\
\makebox[3ex]{$^{ac}$}
\begin{minipage}[t]{14cm}
member of National Technical University of Ukraine, Kyiv, Ukraine\
\end{minipage}\\
\makebox[3ex]{$^{ad}$}
\begin{minipage}[t]{14cm}
now at DESY ATLAS group\
\end{minipage}\\
\makebox[3ex]{$^{ae}$}
\begin{minipage}[t]{14cm}
member of National University of Kyiv - Mohyla Academy, Kyiv, Ukraine\
\end{minipage}\\
\makebox[3ex]{$^{af}$}
\begin{minipage}[t]{14cm}
Alexander von Humboldt Professor; also at DESY and University of Oxford\
\end{minipage}\\
\makebox[3ex]{$^{ag}$}
\begin{minipage}[t]{14cm}
STFC Advanced Fellow\
\end{minipage}\\
\makebox[3ex]{$^{ah}$}
\begin{minipage}[t]{14cm}
now at LNF, Frascati, Italy\
\end{minipage}\\
\makebox[3ex]{$^{ai}$}
\begin{minipage}[t]{14cm}
This material was based on work supported by the
 National Science Foundation, while working at the Foundation.\
\end{minipage}\\
\makebox[3ex]{$^{aj}$}
\begin{minipage}[t]{14cm}
also at Max Planck Institute for Physics, Munich, Germany, External Scientific Member\
\end{minipage}\\
\makebox[3ex]{$^{ak}$}
\begin{minipage}[t]{14cm}
now at Tokyo Metropolitan University, Japan\
\end{minipage}\\
\makebox[3ex]{$^{al}$}
\begin{minipage}[t]{14cm}
now at Nihon Institute of Medical Science, Japan\
\end{minipage}\\
\makebox[3ex]{$^{am}$}
\begin{minipage}[t]{14cm}
now at Osaka University, Osaka, Japan\
\end{minipage}\\
\makebox[3ex]{$^{an}$}
\begin{minipage}[t]{14cm}
also at \L\'{o}d\'{z} University, Poland\
\end{minipage}\\
\makebox[3ex]{$^{ao}$}
\begin{minipage}[t]{14cm}
member of \L\'{o}d\'{z} University, Poland\
\end{minipage}\\
\makebox[3ex]{$^{ap}$}
\begin{minipage}[t]{14cm}
now at Department of Physics, Stockholm University, Stockholm, Sweden\
\end{minipage}\\

}


%% file: DESY-13-054-txt.tex
\section{Introduction}
\label{sec-int}

The measurement of charm production in deep inelastic $ep$ scattering (DIS)
is a powerful tool to study quantum chromodynamics (QCD) and the
proton structure. In leading-order QCD, charm production occurs
through the boson--gluon fusion (BGF) process $\gamma^* g \rightarrow c \bar{c}$,
which  is directly sensitive to the gluon content of the proton. 
Different approaches to the calculation of the heavy-quark contribution
to the proton structure functions are currently used in global analyses of
parton density functions (PDFs)~\cite{Kretzer:2003it,Martin:2009iq,Alekhin:2009ni,Ball:2011mu}. 
Comparisons to measurements of charm production in DIS provide direct tests of
these approaches~\cite{h1zeus}. It has also been shown recently that a combined analysis of charm production
and inclusive DIS data can provide a competitive determination of the
charm-quark mass~\cite{Alekhin:2010sv, h1zeus,Alekhin:2012un}.

Several measurements of charm production in DIS  have been
performed at HERA, exploiting reconstructed $D^{0}$~\cite{epj:c63:2009:2:171-188},
 $D^{\pm}$~\cite{zeus:dplus:2012,epj:c63:2009:2:171-188,Abramowicz:2010aa} and 
$D^{*\pm}$~\cite{pr:d69:012004,pl:b407:402,epj:c12:35,pl:b528:199,epj:c38:447,Aaron:2009jy,Aaron:2011gp}
mesons, semi-leptonic decays~\cite{epjc:65:65-79}, and inclusive lifetime methods~\cite{epj:c45:23,Aaron:2009af} to tag charm. 
In this paper, a new high-statistics measurement of $D^{*\pm}$ 
production via the reaction
$$ e(k) \, p(P) \, \rightarrow \, e'(k') \,
D^{*\pm}(p^{D^*}) \, X \, $$
is presented. The symbols in parenthesis represent the four-momenta
of the incoming ($k$) and outgoing electron ($k'$), of the
incoming proton ($P$), and of the produced $D^{*\pm}$ (${p^{D^*}}$).
The measurement is performed for photon virtualities, $Q^2 \equiv
-{q}^2 = -({k'}-{k})^2$,
in the range $5<Q^2<1000$~GeV$^2$ and for inelasticities, $y \equiv ({P} \cdot
{q}) / ({P}\cdot {k}) $, in the range $0.02<y<0.7$. 

The $D^{*+}$ mesons\footnote{Hereafter the charge conjugated states are implied.} were reconstructed  through the 
decay $D^{*+} \rightarrow D^0\pi^{+}$ with $ D^0 \rightarrow K^{-} \pi^{+}$.
Differential cross sections are presented as a function of
$Q^2$, $y$, the Bjorken-$x$ variable, and of the fraction of the exchanged-photon energy transferred to
the $D^{*+}$ meson in the proton rest frame, $z^{D^*} \equiv ({P} \cdot {p^{D^*}}) / ({P}\cdot {q})$, as well as of the  $D^{*+}$ pseudorapidity, $\eta^{D^*}$, and the transverse momentum, $p_T^{D^*}$, in the laboratory frame
\footnote{
The ZEUS coordinate system is a right-handed Cartesian system, with the $Z$
axis pointing in the proton beam direction, referred to as the ``forward
direction'', and the $X$ axis pointing towards the centre of HERA.
The coordinate origin is at the nominal interaction point. The pseudorapidity is defined 
as $\eta=-\ln\left(\tan\frac{\theta}{2}\right)$,
where the polar angle, $\theta$, is measured with respect to the proton beam
direction.\xspace}.

Double-differential cross sections in $Q^2$ and $y$ are presented
and used to extract the charm contribution to the proton
structure functions in the form of the reduced charm cross section,
$\sigma_{\rm red}^{c\bar{c}}$. Previous measurements and theoretical calculations are compared to the results.

\section{Experimental set-up}
\label{sec-exp}
The measurement was based on $e^{\pm} p$ collisions collected with the
ZEUS detector at HERA in the period 2004--2007 with an electron\footnote{Hereafter ``electron'' refers to
  both electrons and positrons unless otherwise stated.} beam
energy, $E_e$, of $27.5\gev$ and a proton beam energy, $E_p$, of $920\gev$,
corresponding to a centre-of-mass energy $\sqrt{s}=318\gev$. The corresponding
integrated luminosity, ${\cal  L}=363\pm 7$~pb$^{-1}$, is 
four times larger than that used for the previous ZEUS measurement~\cite{pr:d69:012004}.

A detailed description of the ZEUS detector can be found
elsewhere~\cite{zeus:1993:bluebook}. 
In the kinematic range of the analysis, charged particles were tracked
in the central tracking detector (CTD)~\citeCTD and in the microvertex
detector (MVD)~\citeMVD. These components operated in a magnetic
field of $1.43\Tesla$ provided by a thin superconducting solenoid. The
CTD consisted of 72~cylindrical drift chamber layers, organised in
nine superlayers covering the polar-angle region
\mbox{$15^\circ<\theta<164^\circ$}. 
The MVD consisted of a barrel (BMVD) and a forward (FMVD) section with 
three cylindrical layers and four vertical planes of single-sided silicon strip sensors in the BMVD
and FMVD respectively. The BMVD provided polar-angle coverage for
tracks crossing the three layers from $30^\circ$ to $150^\circ$. The
FMVD extended the polar-angle coverage in the forward region down to
$7^\circ$. For CTD--MVD tracks that pass through all nine CTD superlayers, the
momentum resolution was  $\sigma(p_T )/p_T = 0.0029p_T \oplus 0.0081 \oplus 0.0012/p_T$, with $p_T$ in GeV.

The high-resolution uranium--scintillator calorimeter (CAL)~\citeCAL
consisted of three parts: the forward, the barrel, and
the rear (RCAL) calorimeters. Under test-beam conditions, the CAL single-particle relative
energy resolutions were $\sigma(E)/E=0.18/\sqrt{E}$ for electrons and $\sigma(E)/E=0.35/\sqrt{E}$ for
hadrons, with $E$ in $\Gev$. The energy of electrons hitting the RCAL
was corrected for the presence of dead material using
the rear presampler detector~\cite{nim:a382:419} and the
small angle rear tracking detector (SRTD)~\cite{nim:a401:63}.

The luminosity was measured using the Bethe--Heitler reaction $ep
\rightarrow e\gamma p$ by a luminosity detector which consisted of two
independent systems: a lead--scintillator calorimeter
\cite{desy-92-066,*zfp:c63:391,*acpp:b32:2025} and a magnetic
spectrometer~\cite{nim:a565:572}.

\section{QCD calculations}
\label{sec-theo}
Cross sections for heavy-quark production in DIS were calculated at next-to-leading
order (NLO), i.e. $O(\alpha_s^2)$, in the fixed-flavour-number
 scheme (FFNS), in which only light flavours and gluons are present as partons in the proton and
 heavy quarks are produced in the hard interaction~\cite{np:b374:36}.
The program  {\sc Hvqdis}~\cite{np:b452:109,pl:b353:535} was used to
compute single- and double-differential $D^{*+}$ cross sections.

The parameters used as input to {\sc Hvqdis} are listed below,
together with the variations used to evaluate the uncertainty on the
theoretical prediction:
\begin{itemize}
\item charm-quark pole mass: $m_c=1.50 \pm 0.15 \gev$;
\item renormalisation ($\mu_R$) and factorisation ($\mu_F$) scales:  $\mu_R=\mu_F=
  \sqrt{Q^2+4m_c^2}$, varied independently up and down by a factor two;
\item strong coupling constant in the three-flavour FFNS:
  $\alpha_s^{\rm{nf}=3}(M_Z)=0.105\pm 0.002$;
\item the PDFs and their uncertainties, taken from a FFNS variant~\cite{h1zeus} of the
 {\sc HERAPDF1.0} fit~\cite{jhep:2010}. The central PDF set was
 obtained from a fit performed using the same values
 of $m_c$,  $\mu_R$, $\mu_F$ and $\alpha_s$ as used in the {\sc Hvqdis} program. For each 
 variation of these parameters in  {\sc Hvqdis}, a different PDF set was used, in which
 the parameters were varied consistently.
\end{itemize}

The NLO calculation provided differential cross sections for charm 
quarks. The fragmentation model described in a previous
publication~\cite{h1zeus} was used to compare to
the measured $D^{*+}$ cross sections. This model is based on the fragmentation function of Kartvelishvili et
al.~\cite{Kartvelishvili:1977pi},  controlled by the parameter
$\alpha_K$, to describe the fraction of the charm momentum transferred
to the $D^{*+}$ mesons. It also implements a transverse fragmentation
component by assigning to the $D^{*+}$ meson a transverse momentum, $k_T$,
with respect  to the charm-quark direction. The uncertainty on the
fragmentation model was estimated by varying $\alpha_K$ and the average
$k_T$ according to the original prescription~\cite{h1zeus}.
The fraction of charm quarks hadronising into $D^{*+}$ mesons
was set to $f(c\rightarrow D^{*+})=0.2287 \pm 0.0056$~\cite{Lohrmann:2011ce}.

For the inclusive cross section, theoretical predictions were also obtained
in the generalised-mass variable-flavour-number scheme (GM-VFNS). In
this scheme, charm quarks are treated as massive particles for $Q^2\le m_c^2$ and as massless
partons for $Q^2\gg m_c^2$, interpolating in the intermediate region~\cite{Aivazis:1993pi,Buza:1995ie,Collins:1998rz}.
The calculation was performed using the Roberts--Thorne (RT)
``standard''~\cite{pl:b421:303,pr:d57:6871} variant of the GM-VFNS at
NLO, corresponding to $O(\alpha^2_s)$ for the $Q^2\le m_c^2$ part and
to $O(\alpha_s)$ for the $Q^2\gg m_c^2$ part.
PDFs obtained from the HERAPDF1.5\citeHPDF fit
to inclusive HERA data were used. The central prediction
was obtained for $m_c=1.5$~GeV.
To evaluate the theoretical uncertainty, the calculation was
repeated varying the PDF set and its parameters according
to the systematic variations associated with the HERAPDF1.5 fit.
The dominant source of uncertainty was the charm-quark mass, which was varied in the range $1.35<m_c<1.65\gev$.

\section{Monte Carlo samples}
Monte Carlo (MC) samples were used to calculate the experimental
acceptance and to estimate the background contamination.
MC samples of charm and beauty DIS events  were generated using  {\sc Rapgap 3.00}~\cite{cpc:86:147}.
The main sample consisted of events generated according to the LO BGF process. Radiative
 QED corrections to the BGF process  were included through {\sc Heracles 4.6}~\cite{cpc:69:155}. Additional  {\sc Rapgap} samples were generated for diffractive charm production
and for the resolved-photon processes $gg \rightarrow c\bar{c}$ and
$cg\rightarrow cg$, in which one of the incoming partons originates
from the exchanged photon.
Charm photoproduction was simulated using {\sc Pythia 6.2}~\cite{hep-ph-0108264}.

Both  {\sc Rapgap} and {\sc Pythia} use parton showers to simulate
higher-order QCD effects and use the {\sc Pythia/Jetset} hadronisation model~\cite{hep-ph-0108264}. 
All samples were generated using the CTEQ5L~\cite{epj:c12:375} proton
PDFs and, for resolved-photon processes, the GRV-G
LO~\cite{pr:d45:3986} photon PDFs. 
The diffractive samples were generated using  the ``{\sc H1} fit 2''~\cite{pl:b520:191}
diffractive PDFs. The heavy-quark masses were set to $m_c=1.5\gev$
and $m_b=4.75\gev$. Masses, widths and lifetimes of charmed mesons
were taken from  PDG2010~\cite{Nakamura:2010zzi}. 

The MC samples correspond to about four times the luminosity of the data and
were passed through a full simulation of the 
ZEUS detector based on {\sc Geant 3.21}~\cite{tech:cern-dd-ee-84-1}.
They were then subjected to the same trigger criteria and reconstructed 
with the same programs as used for the data.

\section{Event selection and signal extraction}
\label{sub:ev_sel}
\subsection{DIS event selection}
A three-level trigger system was used to select DIS events online
\cite{zeus:1993:bluebook,uproc:chep:1992:222,nim:a580:1257} by
requiring electromagnetic energy deposits in the CAL at the first level and
applying loose DIS selection criteria at the second and third levels.

Offline, the hadronic system was reconstructed using energy-flow objects (EFOs)~\cite{thesis:briskin:1998}
which combine tracking and calorimeter information. The electron was
identified using a neural-network algorithm~\cite{nim:a365:508}.
The kinematical variables $Q^2$, $y$, and $x$ were reconstructed
using the $\Sigma$ method~\cite{nim:a361:197}.  The variable $z^{D^*}$ was
reconstructed according to $z^{D^*}  = (E^{D^*}-p_Z^{D^*})/(2 E_e y_{\mathrm{JB}})$, where
$y_{\mathrm{JB}}$ is
the inelasticity reconstructed with the
Jacquet-Blondel method~\cite{proc:epfacility:1979:391} and $E^{D^*}$ and $p_Z^{D^*}$ are
the $D^{*+}$ energy and longitudinal momentum, respectively.

The following  criteria were applied to select DIS events~\cite{thesis:bachynska:2012}:
\begin{itemize}
\item 
$E_{e'}>10\gev$, where $E_{e'}$ is the energy of the scattered electron;
\item
 $y_{e}<0.7$,  $y_{\mathrm{JB}}>0.02$, where $y_e$ is the inelasticity
 reconstructed from the scattered electron;
\item
$40<\!E-P_{Z}\!<70 \gev$, where  $E-P_{Z}$ is the global difference of
energy and longitudinal momentum, obtained by summing the electron and the hadronic final state,
which is expected to be $2 E_e=55\gev$ for fully contained events;
\item
the $Z$ position of the primary vertex, $Z_{\rm vtx}$, was required to be in the range $|Z_{\rm vtx}| < 30 \, \rm cm$; 
\item
the impact point of the scattered electron on the RCAL was
required to lie outside a square region around the beam-pipe hole:
\mbox{$|X_e|>15$~cm} or \mbox{$|Y_e|>15$~cm};
\item $5 < Q^2 < 1000$ GeV$^2$, where $Q^2$ is reconstructed with the $\Sigma$ method.
\end{itemize} 
\subsection{Selection of ${\mathbf D^{*+}}$ candidates and signal extraction}
\label{sub:ds_sel}

The $D^{*+}$ mesons were identified using the decay channel 
$D^{*+}\to D^0\pi^{+}_s$ with the subsequent decay $D^0\to K^-\pi^+$, where $\pi^+_s$ refers to a low-momentum 
(``slow'') pion accompanying the $D^0$. 

Tracks from the $D^{*+}$ decay products were required 
to have at least one hit in the MVD or in the inner superlayer of the CTD and to reach at least the third superlayer. 
Tracks with opposite charge and with  transverse momentum
$p_T^{K,\pi}>0.4\gev$ were combined in pairs to form $D^0$ candidates.
The track parameters were improved by fitting the two tracks to a
common vertex. Pairs incompatible with coming from the same decay
were removed by requiring a distance of closest approach of the two tracks 
of less than 1~mm, and the $\chi^2$ of the two-track vertex fit smaller 
than 20 for one degree of freedom. The tracks were alternately assigned
 the kaon and pion mass and the invariant mass of the pair, $M(K\pi)$, was calculated. Each additional track, with charge opposite to 
that of the kaon track and a transverse momentum
$p_T^{\pi_s}>0.12\gev$, was assigned the pion mass and  combined with the 
$D^0$ candidate to form a $D^{*+}$ candidate. The $\pi_s$ track was
then fitted to the primary vertex of the event, obtained exploiting the other tracks reconstructed in the event and the constraint
 from the average position of the interaction point~\cite{epj:c63:2009:2:171-188}.
The mass difference $\Delta M \equiv M(K\pi\pi_s) - M(K\pi)$ was
used to extract the $D^{*+}$ signal.
 The $D^{*+}$ candidates were required to have \mbox{$1.80 < M(K\pi) <1.92
 \gev$},    $143.2<\Delta M<147.7 \mev$, $1.5 < p_T^{D^*}<20\gev$ and  \mbox{$|\eta^{ D^*}|<1.5$}.

The distribution of $M(K\pi)$ for $D^{*+}$ candidates, without the requirement
 on $M(K\pi)$, is shown in  Fig.~\ref{f:dzero}. Also shown is the distribution of
wrong-sign (WS) candidates, obtained by combining two tracks with the
same charge. The WS distribution
provides an estimate of combinatorial backgrounds. A clear peak at the
$D^0$ mass is visible in the correct-sign (CS) distribution. The excess of
CS candidates at masses below the $D^0$ peak is due to partly-reconstructed $D^0$
decays, mostly $D^0\rightarrow K^- \pi^+  \pi^0$.

The distribution of  $\Delta M$ for $D^{*+}$ candidates, without
the requirement on $\Delta M$, is shown in Fig.~\ref{f:dstar}. A clear $D^{*+}$ peak is seen. 
 The $D^{*+}$ signal was extracted by subtracting the background
estimate from the number of candidates in the signal window
 $143.2<\Delta M<147.7 \mev$. The background estimate was obtained by
 fitting simultaneously the CS and WS distributions to the parametrisation
\begin{eqnarray*}
\label{eq:granet}
{\rm WS:}  & f_{\rm ws}(\zeta) =& A \, \zeta^B \, e^{-C\zeta}, \\
{\rm CS:}  & f_{\rm cs}(\zeta) = &D \, f_{\rm ws}(\zeta) ,
\end{eqnarray*}
where  $A$, $B$, $C$, $D$ are free parameters of the
fit~\cite{Granet:1977px} and $\zeta=\Delta M - m_{\pi^+}$. The fit was performed in the region $\Delta M<168 \mev$.
The region with a possible signal contribution, $140<\Delta M<150 \mev$,
 was removed from the fit to the CS distribution.
 The parameter $D$, which represents the normalisation of the CS
background with respect to the WS distribution, is slightly larger than unity, $D=1.021\pm0.005$.
This is consistent with the MC estimation of the additional combinatorial background component in the CS
distribution due to real $D^0\rightarrow K\pi$ decays associated with a random track
to form a CS $D^{*+}$ candidate.  The total signal is $N_{\rm data}^{D^{*\pm}}  =12893 \pm 185 $.

The amount of signal lost due to the tails of the $D^0$ mass peak leaking outside the $M(K\pi)$ window  was
estimated by enlarging the mass window to $1.7<M(K\pi)<2.0 \gev$. The fraction of additional
$D^{*+}$ found within the enlarged window was 13\%, including the contribution from partly
reconstructed $D^0$. 
This fraction, as well as its dependence on $p_T^{D^*}$ and $\eta^{D^*}$ and on the width of the $M(K\pi)$ window,
was found to be well reproduced by MC.
 The signal in the tails of the $D^{*+}$ peak outside the
$\Delta M$ window was estimated similarly,
enlarging the signal window to  $140<\Delta M<150 \mev$. The fraction of additional $D^{*+}$
was 6\% on average, with a  dependence on the transverse
momentum of the slow pion,  due to the momentum and
angular resolution degrading at low $p_T^{\pi_S}$.
 This effect is not completely reproduced by the MC. 
An acceptance correction~\cite{thesis:bachynska:2012} dependent on $p_T^{\pi_s}$ was then applied, ranging 
from $\approx 10\%$ at $p_T^{\pi_s}=0.12\gev$ to $\approx 1\%$ at
large $p_T^{\pi_s}$.

\section{Cross-section extraction}
\label{sec:acceptance}
The differential cross sections, $d\sigma_{\rm vis}/d\xi$,
for producing a $D^{*+}$ in the
``visible'' phase space  $1.5<p_T^{D^*}<20 \gev$,   $|\eta^{D^*}|<1.5 $, $5<Q^2<1000 \gev ^2$ and $0.02<y<0.7$
 was obtained as
$$\frac{d\sigma_{\rm vis}}{d\xi} = \frac {N_{\rm data}^{D^*}  -
   N_{\rm \gamma p}^{D^*} }
  {\Delta \xi  \, \cdot {\cal A} \cdot \, BR \, \cdot {\cal L}}
  \cdot C_{r},$$
where $N_{\rm data}^{D^*}$ is the signal extracted in
a bin of a given variable $\xi$, $N_{\rm \gamma p}^{D^*}$ is the photoproduction background, $\Delta \xi$ is the bin
size, ${\cal A}$ is the acceptance, $BR = {\cal B}(D^{*+} \rightarrow D^0
\pi^+) \times  {\cal B}(D^0 \rightarrow K^-
\pi^+) = 0.0263 \pm 0.0004$~\cite{pdg2012} is the branching
ratio, ${\cal L}$ is the integrated luminosity and $C_{r}$
is the QED radiative correction.

The background from charm photoproduction ($Q^2 < 1.5$~GeV$^2$) was
evaluated using the photoproduction MC sample, normalised to
the luminosity using the cross sections previously measured by ZEUS\citeDsPhP.

The acceptance, ${\cal A}$, was calculated as the ratio between the number of reconstructed
and generated $D^{*+}$ in the bin, using a signal MC based on a mix of charm and
beauty production.  The beauty MC was normalised to 1.6  times the 
cross section given by {\sc Rapgap}, consistent with ZEUS 
measurements~\cite{pl:b599:173,epjc:65:65-79,Abramowicz:2010zq,Abramowicz:2011kj}.
The charm MC contained non-diffractive and diffractive components, summed according to the relative cross
sections as given by {\sc Rapgap}. The normalisation of the charm MC was adjusted such that the sum of all the MC components 
reproduced the number of $D^{*+}$ mesons in the data.
Resolved-photon processes were not included. They were only used for systematic checks.
The $\eta^{D^*}$ and $p_T^{D^*}$ distributions of the charm MC were
reweighted~\cite{thesis:bachynska:2012} to improve the agreement with data, with the $p_T^{D^*}$
weights dependent on $Q^2$. 

The acceptance as determined by the MC was corrected to account for
imperfections in the simulation of the trigger and track-reconstruction efficiencies.
One of the main sources of track-reconstruction inefficiency
for charged pions and kaons were hadronic interactions in
the material between the interaction point and the CTD. 
This effect was studied using special tracks from $ep \rightarrow e
\rho^0$ with $\rho^0 \rightarrow \pi^+ \pi^-$
events, reconstructed from MVD hit information alone~\cite{thesis:libov:2012}. For these tracks, an
extension into the CTD was searched for. In addition, the $p_T$
dependence of the tracking efficiency  was studied by exploiting the isotropic angular distribution of pions
from $K_S^0$ decays. The studies showed that the MC slightly underestimated the
effect of nuclear interactions. For central pions with  $p_T \approx 1\gev$,
the track-reconstruction inefficiency due to hadronic interactions was measured to be $(7 \pm 1)$\% while the MC predicted
5\%. The  track-efficiency correction was applied as a function
of $\eta$ and $p_T$ of each track. For $p_{T}>1.5\gev$, no correction was necessary. 

The acceptance ranges
from ${\cal A} \approx 10\%$ in the lowest $p_T^{D^*}$ and $Q^2$ bins to
${\cal A} \approx 45\%$ in the highest $p_T^{D^*}$ and $Q^2$ bins.
Fig.~\ref{f:control} shows $N_{\rm data}^{D^*} / \Delta \xi$ for $\xi=p_{T}^{D^*},\,\eta^{D^*}\,,Q^2,\,y$ and $z^{D^*}$.
The sum of the different MC samples is compared to the data.  The agreement is satisfactory.

The cross sections were corrected to the QED Born level,
using a running coupling constant $\alpha_{\rm em}(Q^2)$,
such that they can be compared directly to the QCD predictions from the {\sc Hvqdis} program.
The radiative corrections were obtained as
 $C_{r}=\sigma^{\rm Born}_{\rm vis}/\sigma^{\rm rad}_{\rm vis}$, where
 $\sigma^{\rm Born}_{\rm vis}$ is the
 {\sc Rapgap} cross section with the QED corrections turned off but 
keeping $\alpha_{\rm em}$ running and $\sigma^{\rm rad}_{\rm vis}$ is
the {\sc Rapgap} cross section with the full QED corrections, as in the standard
MC samples. 

\section{Systematic uncertainties}
\label{sub:syst}
The experimental systematic uncertainties are listed
below~\cite{thesis:bachynska:2012}, with their typical effect on the measured cross sections is given in parenthesis:
  \begin{enumerate}
    \item[$\delta_{1}$] energy-scale uncertainty on the hadronic
      system  of $\pm 2\%$   ($\pm 1\%$,  up to $\pm 10\%$ at low $y$);
    \item[$\delta_{2}$] electron energy-scale uncertainty of $\pm 1\%$~\cite{thesis:shimizu:2009}
     ($\pm 1\%$, up to $\pm 7\%$ at low $y$);
    \item[$\delta_{3}$] alignment uncertainty on the electron impact
      point on the RCAL, estimated by varying the cut on the
      electron position in the MC by $\pm2$~mm separately for the $X_e$ and $Y_e$
      coordinates ~\cite{thesis:shimizu:2009}
      ($\pm7\%$ at low $Q^2$ and low $y$,  negligible at large $Q^2$);
    \item[$\delta_{4}$] uncertainty on the position of the electron
      impact point on the RCAL due to imperfections in the simulation
      of the shower shape and of the detector resolution, estimated by
      loosening the cut on the electron position by 1~cm
      ($|X_e|>14$~cm or $|Y_e|>14$~cm) both in data and in MC (up to $\pm 10\%$ at low $y$ and low
      $Q^2$, negligible at large $Q^2$);
    \item [$\delta_{5}$] uncertainty on the background shape in
      $\Delta M$,  estimated by replacing the function $f_{\rm cs}(\zeta)$ by
      $f_{\rm cs}^{\prime}(\zeta)=A \zeta^{\frac{2}{3}}+ B \zeta +C
      \zeta^{\frac{1}{2}}$  ($+0.3\%$);
    \item[$\delta_{6}$] a further uncertainty on the background shape, evaluated by reducing the fit range from  $\Delta M< 168\mev$
      to $\Delta M < 165 \mev$ ($+0.5\%$); 
    \item[$\delta_{7}$]  uncertainty on the amount of signal outside the $\Delta M$ window, evaluated by varying the $p_T^{\pi_S}$-dependent correction by its
      uncertainty  ($\pm 1.5\%$, up to $\pm 3\%$ at low $p_T$);
    \item[$\delta_{8}$]  uncertainty on the amount of signal outside the $M(K\pi)$ window, estimated by comparing data and MC in an
      enlarged mass range  ($+2\%$);
     \item[$\delta_{9}$] uncertainty on the track-efficiency, evaluated by varying the
      track efficiency correction applied to MC by the associated
      uncertainty ($\pm 2\%$); 
     \item[$\delta_{10}$]  uncertainty on the trigger efficiency,
       evaluated using independent triggers ($\pm 0.5\%$);
     \item[$\delta_{11}$] statistical uncertainty on the calculation
       of the acceptance   ($\pm 1\%$);
    \item[$\delta_{12}$] uncertainty on the normalisation of the beauty MC of $\pm 50\%$  to cover the range allowed
 by ZEUS measurements~\cite{pl:b599:173,Abramowicz:2011kj} ($\pm 0.3\%$);
    \item[$\delta_{13}$] uncertainty on the normalisation of the
      photoproduction MC of $\pm 100\%$
 (up to $\pm3\%$ at high $y$, but negligible elsewhere);
    \item[$\delta_{14}$] uncertainty on the normalisation of the
      diffractive charm MC of $\pm 50\%$ to cover the range allowed by data--MC comparison
 and by previous ZEUS results~\cite{np:b672:3} (up to $\pm4.5\%$ at low $y$, but negligible elsewhere);
    \item[$\delta_{15}$] uncertainty due to the resolved-photon component, evaluated by adding the resolved-photon samples to the charm
      MC normalised according to the generator cross section ($+2\%$);
    \item[$\delta_{16}$] uncertainty on the MC reweighting as a function of
      $p_{T}^{D^*}$ and $ Q^{2 }$, which was varied by $\pm 50\%$ ($\pm2\%$);
    \item[$\delta_{17}$] uncertainty on the MC reweighting as a function of
      $\eta^{D^{*}}$ which was
      replaced by a MC reweighting as a function of  $y$   (from
      $-2\%$  to $+3\%$, depending on $y$);
   \item[$\delta_{18}$] uncertainty on the integrated luminosity of $\pm1.9\%$;
   \item[$\delta_{19}$] uncertainty on the branching ratio $BR$ of $\pm 1.5\%$.
   \end{enumerate}
All the systematic  uncertainties, except the overall normalisations $\delta_{18}$
and $\delta_{19}$, were added in quadrature to the statistical
uncertainties to obtain the total error bars in the figures.

\section{Results}
\label{sec:res}
Single- and double-differential cross sections have been measured in the phase space
$$ 5<Q^2<1000 \gev^2  ;~   0.02<y<0.7 ;~   1.5<p_T^{D^*}<20 \gev   ;~   |\eta^{D^*}|<1.5 . $$
Differential cross sections in $p_T^{D^*}$,  $\eta^{D^*}$ and
$z^{D^*}$ are reported in Tables~\ref{tab:xs_pt}--\ref{tab:xs_z} and in
Fig.~\ref{f:single_2}.  The cross section decreases steeply with
$p_T^{D^*}$ and is almost constant in $\eta^{D^*}$. 
The  NLO calculations based on {\sc Hvqdis}
and the {\sc Rapgap} MC implementing the leading-order BGF process
are compared to the data. As the {\sc Rapgap} MC is based on leading-order matrix elements, it
is not expected to estimate the normalisation correctly. Therefore the {\sc Rapgap} prediction was normalised to the data, scaling it by 1.1, to allow a direct comparison of the
shapes. The data are well described by the NLO calculation and by
{\sc Rapgap} with the exception of the shape in $z^{D^*}$, which
is not well reproduced by the NLO calculation, suggesting possible
imperfections in the fragmentation model.

Differential cross sections in  $Q^2$, $y$ and $x$ are
reported in Tables~\ref{tab:xs_q2}--\ref{tab:xs_x} and in
Fig.~\ref{f:single_1}. The results are reasonably well described by the NLO
calculation.  The MC predictions reproduce the shapes of the data, except
for the high-$Q^2$ tail, where the MC prediction  is too high,
and for $d\sigma/dy$, where the prediction is too low at low $y$ and too high
at large $y$. These imperfections in the MC are to be expected in the absence of
higher-order terms in {\sc Rapgap}.

Visible cross sections in two-dimensional bins of $Q^2$ and $y$, $\sigma_{\rm vis}$, are given in Table~\ref{tab:2d_xs}.
The corresponding bin-averaged double-differential cross sections are shown in Figs.~\ref{f:double_1} and \ref{f:double_2}.
The values of the individual systematic uncertainties on the double-differential cross sections are given in Table~\ref{t:syst}. 
Measurements  performed in the same phase space by the H1
Collaboration~\cite{Aaron:2011gp, Aaron:2009jy}, which are the most
precise previous measurement of $D^{*+}$ production in DIS, are
compared to the present results.  The two data sets are in agreement and have similar precision.
The double-differential cross sections are well described by the NLO
calculation.

In a previous ZEUS measurement~\cite{pr:d69:012004}, a possible excess in the $D^{*+}$ yield in $e^-p$ collisions was
observed with respect to $e^+p$ collisions. 
The ratio of observed rates, increasing with $Q^2$, was $r^{e^-p}/r^{e^+p} = 1.67 \pm
0.21 ({\rm stat.})$ for $40<Q^2<1000\gev^2$. 
 The measurement was based on a luminosity of
$17$ ($65$)~pb$^{-1}$ of $e^-p$ ($e^+p$) collisions. The present measurement is
based on an independent data set, consisting of $187$ ($174$)~pb$^{-1}$ of
$e^-p$ ($e^+p$) collisions. Fig.~\ref{f:epm} shows the cross-section
ratio as a function of $Q^2$. Only statistical uncertainties are shown
since systematic effects mostly cancel in the ratio. No deviation from unity is observed,
confirming the original interpretation of the $e^-p$ excess as a
statistical fluctuation.

\section{Charm reduced cross sections}
\label{sec:red}
The reduced cross section for charm, $\sigma^{c\bar{c}}_{\rm red}$,  and 
the charm contribution to the proton structure functions, $F_2^{c\bar{c}}$ and $F_L^{c\bar{c}}$, are defined as:
$$\frac{d^2\sigma^{c\bar{c}}}{dx\, dQ^2}
 = \frac{2\pi \alpha_{em}^2}{x Q^4}  ~Y_+ ~ \sigma^{c\bar{c}}_{\rm red}(x,Q^2,s),$$
$$ \sigma^{c\bar{c}}_{\rm red}(x,Q^2,s)  =   F_2^{c\bar{c}}(x,Q^2) - \frac{y^2}{Y_+} F_L^{c\bar{c}} (x,Q^2),$$
where $Y_+=1+(1-y)^2$. 

The {\sc Hvqdis} program was used to extrapolate the measured visible $D^{*+}$
cross sections in bins of $y$ and $Q^2$, $\sigma_{\rm vis}$, to the full phase space:
$$ \sigma^{c\bar{c}}_{\rm red}(x, Q^2) = 
   \left( \sigma_{\rm vis}  -  \sigma_{\rm   vis}^{\rm beauty}
   \right) \left(  \frac{\sigma^{c\bar{c}}_{\rm red, \, H\textsc{vqdis}  }(x, Q^2)}{\sigma_{\rm
    vis, \,  H\textsc{vqdis}  } }\right), $$
where $\sigma_{\rm   vis}^{\rm beauty}$ is the beauty contribution as
predicted by the {\sc Rapgap} MC, normalised as discussed in Section~\ref{sec:acceptance}, and 
$\sigma^{c\bar{c}}_{\rm red, \, H\textsc{vqdis} }$, $ \sigma_{\rm vis, \,  H\textsc{vqdis} } $ are the charm reduced 
and the visible $D^{*+}$ cross sections, respectively, as given by {\sc Hvqdis}.
 The reference values of $x$ and $Q^2$ were chosen close to the average $x$ and $Q^2$ of
the bins.  The kinematic acceptance of the visible phase space, defined as
${\cal  A}_{ps}=\sigma_{\rm vis}/(\sigma^{c\bar{c}} \cdot 2\,f(c\rightarrow
D^{*+}) )$, where $\sigma^{c\bar{c}}$  is the charm production total cross section in the $y$ and $Q^2$ bin,
 ranges from $17\%$ to $64\%$, depending on the bin.

Following the method used in the previously published combination of ZEUS and H1 results~\cite{h1zeus}, 
the {\sc Hvqdis} and fragmentation variations described in Section~\ref{sec-theo}
were used to determine the theoretical uncertainty on the extraction
of $\sigma^{c\bar{c}}_{\rm red}$. The scales $\mu_R$ and $\mu_F$ were varied simultaneously 
rather than independently as in the theoretical uncertainty for the
differential cross sections. An additional uncertainty originates from
the subtracted beauty component that was varied by $\pm50\%$.
The theoretical uncertainties due the extrapolation on $\sigma_{\rm red}^{c\bar{c}}(x,Q^2)$ are given in Table~\ref{t:theo}.
The experimental part of the uncertainties on $\sigma_{\rm red}^{c\bar{c}}(x,Q^2)$ is defined as the quadratic sum of the statistical
 and the experimental systematic uncertainties described in Section~\ref{sub:syst}.

The results are reported in Table~\ref{t:sig_red} and are shown in Fig.~\ref{f:sigred_1}. The combined
result based on previous H1 and ZEUS charm measurements~\cite{h1zeus} and a recent ZEUS measurement with $D^+$
 mesons~\cite{zeus:dplus:2012}, not included in the combined results, are also shown.
All three measurements are in good agreement. 
The $D^*$ measurement has a precision close to that of the combined result in some parts of the phase space.
 The GM-VFNS calculation, based on the HERAPDF1.5
parton-density fit to inclusive HERA data, is compared to the present measurement and shown in Fig.~\ref{f:sigred_2}. 
The uncertainty on the prediction is dominated by the charm-quark mass.
The prediction is in good agreement with the data.

\section{Conclusions}
\label{sec-conc}
Differential cross sections for the production of $D^{*\pm}$ mesons in DIS have been measured
with the ZEUS detector in the kinematic range
$$ 5<Q^2<1000 \gev^2  ;~   0.02<y<0.7 ;~   1.5<p_T^{D^*}<20 \gev   ;~
|\eta^{D^*}|<1.5, $$
using data from an integrated luminosity of $363$~pb$^{-1}$.
The new data represents one of the most precise measurements of charm production in DIS obtained to date.
The data are reasonably well described by NLO QCD calculations and  are in agreement with previously published results.

The measurements have been used to extract the reduced cross
sections for charm  $\sigma_{\rm red}^{c\bar{c}}$. A GM-VFNS calculation based on a
PDF fit to inclusive DIS HERA data agrees well with the results. This demonstrates 
a consistent description of charm and inclusive data within the NLO QCD
framework. 

\section*{Acknowledgements}
\label{sec-ack}

\Zacknowledge

\vfill\eject

%% file: DESY-13-054-ref.tex
{
\ifzeusbst
  \bibliographystyle{../BiBTeX/bst/l4z_default}
\fi
\ifzdrftbst
  \bibliographystyle{../BiBTeX/bst/l4z_draft}
\fi
\ifzbstepj
  \bibliographystyle{../BiBTeX/user/l4z_epj-ICB}
\fi
\ifzbstnp
  \bibliographystyle{../BiBTeX/bst/l4z_np}
\fi
\ifzbstpl
  \bibliographystyle{../BiBTeX/bst/l4z_pl}
\fi
{\raggedright
\bibliography{../BiBTeX/user/syn,%
              ../BiBTeX/user/myref,%
              ../BiBTeX/bib/l4z_zeus,%
              ../BiBTeX/bib/l4z_h1,%
              ../BiBTeX/bib/l4z_articles,%
              ../BiBTeX/bib/l4z_books,%
              ../BiBTeX/bib/l4z_conferences,%
              ../BiBTeX/bib/l4z_misc,%
              ../BiBTeX/bib/l4z_preprints,%
              ../BiBTeX/bib/l4z_temporary}}      
}
\vfill\eject

%% file: DESY-13-054-tab.tex
\begin{table}[t]
\begin{center}
\tabcolsep2.1mm
\renewcommand*{\arraystretch}{1.2}
\begin{tabular}{|c|c|c|c|c|}
\hline
$p_{T}^{D^*}$ &$ \frac{d\sigma}{d p_{T}^{D^{*}}}$ &$ \delta_{\rm stat}$  &$\delta_{\rm syst}$& $C_{r}$  \\
  (\gev)       & (nb/\gev)                         &$(\%)$            &$(\%)$ & \\ 
\hline
1.50 : 1.88&    2.16&    9.9&    $ ^{+ 7.0}_ {-5.5} $    &1.03\\
\hline
1.88 : 2.28&    2.30&    5.8&    $ ^{+ 5.4}_ {-5.8} $    &1.04\\
\hline
2.28 : 2.68&    1.95&    4.4&    $ ^{+ 5.0}_ {-4.4} $    &1.03\\
\hline
2.68 : 3.08&    1.63&    4.0&    $ ^{+ 4.7}_ {-4.0} $    &1.03\\
\hline
3.08 : 3.50&    1.22&    3.8&    $ ^{+ 4.9}_ {-4.2} $    &1.04\\
\hline
3.50 : 4.00&    9.71$\times 10^{-1}$&     3.4&    $ ^{+ 4.4}_ {-3.7} $    &1.03\\
\hline
4.00 : 4.75&    6.26$\times 10^{-1}$&     3.2&    $ ^{+ 4.2}_ {-3.5}$     &1.05\\
\hline
4.75 : 6.00&    3.32$\times 10^{-1}$&     3.0&    $ ^{+ 4.3}_ {-3.7}$     &1.01\\
\hline
6.00 : 8.00&    1.21$\times 10^{-1}$&     4.1&    $ ^{+ 4.1}_ {-3.8} $    &1.06\\
\hline
8.00 : 11.00&   3.31$\times 10^{-2}$&      6.0&    $ ^{+ 4.4}_ {-3.7} $    &1.11\\
\hline
11.00 : 20.00&        3.60$\times 10^{-3}$&     12&   $ ^{+ 5.3}_ {-6.1} $    &1.11\\
\hline
\end{tabular}
\caption{Differential cross section for $D^{*\pm}$ production in $
  p_{T}^{D^*}$, in the kinematic range $ 5<Q^2<1000 \gev^2$,
  $0.02<y<0.7$, $1.5<p_T^{D^*}<20 \gev$, $|\eta^{D^*}|<1.5 $.
 The columns list the bin range, the bin-averaged differential cross
 section, the statistical, $\delta_{stat}$, and systematic, $\delta_{syst}$, uncertainties, and the QED correction factors, $C_r$. The overall normalization uncertainties from the luminosity
 ($\pm 1.9$\%) and branching ratio of the $D^{*\pm}$ decay channel ($\pm1.5$\%) are not included.}
\label{tab:xs_pt}
\end{center}
\end{table}

\begin{table}[t]
\begin{center}
\tabcolsep2.1mm
\renewcommand*{\arraystretch}{1.2}
\begin{tabular}{|c|c|c|c|c|}
 \hline
$\eta^{D^{*}}$    &$ \frac{d\sigma}{d \eta^{D^{*}}} $&$ \delta_{\rm stat}$ &$\delta_{\rm syst}$ & $C_{r}$  \\
                  & (nb)                          & $(\%)$          &  $(\%)$        &          \\
\hline
$-1.50 : -1.25$&      1.48&    7.5&    $ ^{+ 6.8}_ {-6.7} $   &1.06\\
\hline
$-1.25 : -1.00$&      1.66&    5.4&    $ ^{+ 5.6}_ {-5.3} $   &1.05\\
\hline
$-1.00 : -0.75$&      1.61&    4.9&    $ ^{+ 6.1}_ {-4.4} $   &1.05\\
\hline
$-0.75 : -0.50$&       1.85&    4.2&    $ ^{+ 4.6}_ {-3.8} $   &1.03\\
\hline
$-0.50 : -0.25$&       1.94&    4.2&    $ ^{+ 4.3}_ {-3.5} $   &1.03\\
\hline
$-0.25 :~~ 0.00$&       2.02&    4.0&    $ ^{+ 4.3}_ {-3.7} $   &1.04\\
\hline
$0.00 : 0.25$&        1.90&    4.4&    $ ^{+ 4.2}_ {-3.4} $   &1.04\\
\hline
$0.25 : 0.50$&        1.97&    4.4&    $ ^{+ 4.3}_ {-3.3} $   &1.05\\
\hline
$0.50 : 0.75$&        1.96&    4.7&    $ ^{+ 4.5}_ {-3.6} $   &1.03\\
\hline
$0.75 : 1.00$&        2.02&    4.9&    $ ^{+ 4.8}_ {-4.2} $   &1.02\\
\hline
$1.00 : 1.25$&        2.00&    5.8&    $ ^{+ 5.3}_ {-5.1} $   &1.01\\
\hline
$1.25 : 1.50$&        1.84&    7.7&    $ ^{+ 7.4}_ {-5.6} $   &1.01\\
\hline
\end{tabular}
\caption{Differential cross section of $D^{*\pm}$ production in $
  \eta^{D^*}$. See Table~\ref{tab:xs_pt} for other details.}
\label{tab:xs_eta}
\end{center}
\end{table}

\begin{table}[t]
\begin{center}
\tabcolsep2.1mm
\renewcommand*{\arraystretch}{1.2}
\begin{tabular}{|c|c|c|c|c|}
 \hline
$z^{D^{*}}$    &$ \frac{d\sigma}{d z^{D^{*}}} $&$ \delta_{\rm stat}$    &$\delta_{\rm syst}$ & $C_{r}$ \\
               &  (nb)                     &  $(\%)$            & $(\%)$              &   \\
\hline
0.000 : 0.100&    3.02&    12&    $ ^{+ 8.6}_ {-7.1} $   & 1.00\\
\hline
0.100 : 0.200&    6.83&    6.1&    $ ^{+ 6.1}_ {-5.0} $    & 1.01\\
\hline
0.200 : 0.325&    8.18&    3.5&    $ ^{+ 5.5}_ {-4.9} $    & 1.02\\
\hline
0.325 : 0.450&    9.20&    2.5&    $ ^{+ 4.6}_ {-3.8} $    & 1.03\\
\hline
0.450 : 0.575&    9.14&    2.3&    $ ^{+ 4.6}_ {-4.0} $    & 1.05\\
\hline
0.575 : 0.800&    5.12&    2.4&    $ ^{+ 6.5}_ {-5.1} $    & 1.07\\
\hline
0.800 : 1.000&    0.063&   9.1&    $ ^{+ 9.9}_ {-8.5} $    & 1.07\\
\hline
\end{tabular}
\caption{Differential cross section of $D^{*\pm}$ production in $
  z^{D^{*}}$.  See Table~\ref{tab:xs_pt} for other details.}
\label{tab:xs_z}
\end{center}
\end{table}


\begin{table}[t]
\begin{center}
\tabcolsep1.5mm
\renewcommand*{\arraystretch}{1.2}
\begin{tabular}{|c|c|c|c|c|}
 \hline
$Q^{2}$             &$ \frac{d\sigma}{d Q^{2}} $ &$ \delta_{\rm stat}$  &$\delta_{\rm syst}$ &$C_{r} $  \\
(\gev$^{2}$)        & (nb/\gev$^{2}$)            &$(\%)$            &$(\%)$          &  \\
\hline
5.0 : 8.0&      0.499&     3.9&    $ ^{+ 6.7}_ {-6.1} $    &1.03\\
\hline
8.0 : 10.0&     0.307&     4.3&    $ ^{+ 6.0}_ {-5.2} $    &1.03\\
\hline
10.0 : 13.0&    0.222&     4.0&    $ ^{+ 4.9}_ {-4.1} $    &1.02\\
\hline
13.0 : 19.0&    0.125&     3.5&    $ ^{+ 5.6}_ {-5.0} $    &1.03\\
\hline
19.0 : 27.5&    0.752$\times 10^{-1}$&      3.7&    $ ^{+ 4.9}_ {-4.0} $    &1.04\\
\hline
27.5 : 40.0&    0.415$\times 10^{-1}$&      3.9&    $ ^{+ 4.8}_ {-3.8} $    &1.04\\
\hline
40.0 : 60.0&    0.169$\times 10^{-1}$&      4.7&    $ ^{+ 5.6}_ {-5.6} $    &1.05\\
\hline
60.0 : 100.0&   0.747$\times 10^{-2}$&       5.0&    $ ^{+ 7.1}_ {-5.1} $    &1.06\\
\hline
100.0 : 200.0&  0.171$\times 10^{-2}$&       7.8&    $ ^{+ 6.6}_ {-4.4} $    &1.07\\
\hline
200.0 : 1000.0&   0.140$\times 10^{-3}$&    13&    $ ^{+ 6.1}_ {-5.2} $    &1.14\\
\hline
\end{tabular}
\caption{Differential cross section of $D^{*\pm}$ production in $ Q^{2}$. See
Table~\ref{tab:xs_pt} for other details.}
\label{tab:xs_q2}
\end{center}
\end{table}

\begin{table}[t]
\begin{center}
\tabcolsep2.1mm
\renewcommand*{\arraystretch}{1.2}
 \begin{tabular}{|c|c|c|c|c|}
 \hline
$y$    &$ \frac{d\sigma}{d y} $&$ \delta_{\rm stat}$    &$\delta_{\rm syst}$  &$C_{r}$  \\
       & (nb)                  &$(\%)$                   &$(\%)$                &  \\
\hline
0.02 : 0.05&    1.20 $\times 10^1$&    7.9&    $ ^{+ 16}_ {-12} $ &1.07\\
\hline
0.05 : 0.09&    2.07 $\times 10^1$&    3.4&    $ ^{+ 6.7}_ {-6.5} $   &1.05\\
\hline
0.09 : 0.13&    1.79 $\times 10^1$&    3.4&    $ ^{+ 4.5}_ {-4.0} $   &1.04\\
\hline
0.13 : 0.18&    1.37 $\times 10^1$&    3.6&    $ ^{+ 4.6}_ {-4.8} $   &1.04\\
\hline
0.18 : 0.26&    1.13 $\times 10^1$&    3.3&    $ ^{+ 4.8}_ {-3.7} $   &1.04\\
\hline
0.26 : 0.36&    8.03              &    3.7&    $ ^{+ 4.8}_ {-4.0} $    &1.03\\
\hline
0.36 : 0.50&    5.09&    4.2&    $ ^{+ 5.2}_ {-4.5} $    &1.02\\
\hline
0.50 : 0.70&    2.90&    6.0&    $ ^{+ 9.3}_ {-7.1} $    &1.01\\
\hline
\end{tabular}
\end{center}
\caption{Differential cross section of $D^{*\pm}$ production in $ y$. See
Table~\ref{tab:xs_pt} for other details.}
\label{tab:xs_y}
\end{table}

%
 \begin{table}[t]
\begin{center}
\tabcolsep2.1mm
\renewcommand*{\arraystretch}{1.2}
 \begin{tabular}{|c|c|c|c|c|}
 \hline
$x$  &$ \frac{d\sigma}{d x} $&$ \delta_{\rm stat}$    &$\delta_{\rm syst} $ &$C_{r}$  \\
     &(nb)                   &$(\%)$              &$(\%)$           &  \\
\hline
(0.8 : 4.0) $\times 10 ^{-4}$&  0.475 $\times 10 ^4$&    3.5&    $ ^{+ 6.0}_ {-5.3} $ &1.06\\
\hline
(0.4 : 1.6) $\times 10 ^{-3}$&  0.198 $\times 10 ^4$&    2.1&    $ ^{+ 4.8}_ {-3.9} $ &1.03\\
\hline
(1.6 : 5.0) $\times 10 ^{-3}$&  0.357 $\times 10 ^3$&     2.6&    $ ^{+ 4.9}_ {-3.9} $ &1.02\\
\hline
(0.5 : 1.0) $\times 10 ^{-2}$&  0.553 $\times 10 ^{2}$&      5.7&    $ ^{+ 6.3}_ {-5.1} $ &0.99\\
\hline
(0.1 : 1.0) $\times 10 ^{-1}$&  0.159 $\times 10 ^{1}$ &       10.7&    $ ^{+ 9.2}_ {-8.4} $&1.08\\
\hline
\end{tabular}
\caption{Differential cross section of $D^{*\pm}$ production in $ x$. See
Table~\ref{tab:xs_pt} for other details.}
\label{tab:xs_x}
\end{center}
\end{table}
\begin{table}[t]
\begin{center}
\begin{tabular}{|c|c|r|r|r|r|r|}
 \hline
$Q^{2}$    &$y$   &$ \sigma_{\rm vis}$&$ \delta_{\rm stat}$ &$\delta_{\rm syst}$  & $\sigma_{\rm vis}^{\rm beauty}$ & $C_{r}$ \\
 (\gev$^{2}$)   &   &(pb)             &$(\%)$    &$(\%)$  & (pb) &  \\
\hline
\multirow{5}{*}{5 : 9 } &0.020 : 0.050&    120&    23&    $ ^{+19}_ {- 20} $&0.0  &1.04\\
&0.050 : 0.090&    279&    10&   $ ^{+11}_ {- 11} $&1.5  &1.04\\
&0.090 : 0.160&    421&    6.0&    $ ^{+6.8}_ {- 7.0} $&5.2     &1.04\\
&0.160 : 0.320&    550&    5.3&    $ ^{+6.5}_ {- 5.8} $&11.0    &1.03\\
&0.320 : 0.700&    456&    6.8&    $ ^{+6.3}_ {- 5.5} $&18.2    &1.02\\
\hline
\multirow{5}{*}{9 : 14} &0.020 : 0.050&    108&    14&    $ ^{+17}_ {- 12} $&0.1      &1.05\\
&0.050 : 0.090&    178&    6.5&    $ ^{+7.0}_ {- 6.0} $&1.2      &1.04\\
&0.090 : 0.160&    220&    5.8&    $ ^{+4.7}_ {- 4.6} $&2.9      &1.03\\
&0.160 : 0.320&    352&    5.1&    $ ^{+4.5}_ {- 3.7} $&8.1      &1.02\\
&0.320 : 0.700&    307&    7.2&    $ ^{+6.6}_ {- 5.0} $&12.5      &1.00\\
\hline
\multirow{5}{*}{14 : 23} &0.020 : 0.050&    65.1&    15&    $ ^{+16}_ {- 12} $&0.2      &1.07\\
&0.050 : 0.090&    160&    6.4&    $ ^{+6.2}_ {- 7.2} $&1.2      &1.04\\
&0.090 : 0.160&    205&    5.6&    $ ^{+4.7}_ {- 4.7} $&3.1      &1.03\\
&0.160 : 0.320&    267&    5.9&    $ ^{+4.9}_ {- 4.4} $&9.0      &1.03\\
&0.320 : 0.700&    250&    7.4&    $ ^{+5.7}_ {- 6.7} $&13.5      &1.01\\
\hline
\multirow{5}{*}{23 : 45} &0.020 : 0.050&    37.1&    29&    $ ^{+18}_ {-18} $&0.1      &1.08\\
&0.050 : 0.090&    134&    7.0&    $ ^{+7.5}_ { -7.8}$ &0.9      &1.06\\
&0.090 : 0.160&    196&    5.3&    $ ^{+4.4}_ { -4.3}$ &3.6      &1.05\\
&0.160 : 0.320&    275&    5.1&    $ ^{+4.1}_ { -3.4}$ &10.2      &1.03\\
&0.320 : 0.700&    284&    6.1&    $ ^{+6.4}_ { -4.5}$ &14.7      &1.02\\
\hline
\multirow{5}{*}{45 : 100} &0.020 : 0.050&    14.2&    38&    $ ^{+35}_ {- 18} $&0.0      &1.25\\
&0.050 : 0.090&    72.1&    9.6&    $ ^{+8.0}_ {-7.2} $&1.2      &1.07\\
&0.090 : 0.160&    87.0&    8.4&    $ ^{+4.9}_ {-4.6} $&3.9      &1.04\\
&0.160 : 0.320&    182&    5.7&    $ ^{+5.3}_ {-3.9} $&9.4     &1.04\\
&0.320 : 0.700&    175&    7.6&    $ ^{+6.6}_ {-5.6} $&14.0    &1.02\\
\hline
\multirow{3}{*}{100 : 158} &0.020 : 0.350&    80.2&    11&    $ ^{+7.6}_ {- 4.2} $&5.8      &1.1\\
&0.350 : 0.700&    45.1&    16&    $ ^{+7.6}_ {- 7.8} $&5.0      &0.99\\
\hline
\multirow{3}{*}{158 : 251} &0.020 : 0.300&    49.8&    14&    $ ^{+4.8}_ {- 6.3} $&3.5      &1.16\\
&0.300 : 0.700&    37.3&    17&    $ ^{+6.6}_ {- 4.9} $&4.3      &1.04\\
\hline
\multirow{3}{*}{251 : 1000} &0.020 : 0.275&    28.4&    24&    $ ^{+8.2}_ {- 10} $&2.4     &1.26\\
&0.275 : 0.700&    46.7&    21&    $ ^{+8.7}_ {- 5.1} $&6.9      &1.07\\
\hline
\end{tabular}
\caption{Visible cross sections, $\sigma_{\rm vis}$, for $D^{*\pm}$ production in bins of
  $Q^{2}$ and $y$. The second but last column reports the estimated contribution from beauty decays,
  based on the {\sc Rapgap} beauty MC rescaled to ZEUS data. See
  Table~\ref{tab:xs_pt} for other details.}
\label{tab:2d_xs}
\end{center}
\end{table}

\begin{table}[htb]
\begin{center}
\tiny
 \tabcolsep1.1mm
\renewcommand*{\arraystretch}{1.85}
 \begin{tabular}{|c|c|c|c|c|c|c|c|c|c|c|c|c|c|c|c|c|}
 \hline
$Q^{2}$ (\gev$^{2}$)   &$y$   &$\delta_{1}$&$\delta_{2}$&$\delta_{3}$&$\delta_{4}$&$\delta_{5}$&$\delta_{6}$ &$\delta_{7}$&$\delta_{9}$&$\delta_{11}$&$\delta_{12}$&$\delta_{13}$&$\delta_{14}$ &$\delta_{15}$&$\delta_{16}$ &$\delta_{17}$ \\
   &   & ($\%$) & ($\%$) & ($\%$) & ($\%$) & ($\%$) & ($\%$) & ($\%$) &($\%$) & ($\%$) & ($\%$) &($\%$) &($\%$) &($\%$) &($\%$) &($\%$) \\
\hline
\multirow{5}{*}{5 : 9 } &0.020 : 0.050&    $ _{-12} ^ {+11} $&  $ _{+0.6} ^ {+1.8} $&  $ _{-2.7} ^ {+7.4} $&  $ _{-9.0} ^ {+9.0} $&   -7.0 &   -3.4 &  $ _{-2.4} ^ {+2.5} $&    $ _{-2.5} ^ {+2.5} $&    $ _{-8.8} ^ {+8.8} $&    $ _{-0.0} ^ {+0.0} $&    $ _{+0.0} ^ {+0.0} $&    $ _{+4.2} ^ {-4.2} $&     +1.8 &    $ _{-0.1} ^ {+0.1} $&    +2.2  \\
&0.050 : 0.090&    $  _{-4.0} ^ {+3.5} $&    $  _{-4.8} ^ {+5.1} $&    $  _{-3.0} ^ {+2.3} $&    $  _{-6.6} ^ {+6.6} $&    +0.7  &-1.9  &$  _{-2.2} ^ {+2.4} $&    $  _{-2.1} ^ {+2.1} $&    $  _{-3.4} ^ {+3.4} $&    $  _{+0.1} ^ {-0.1} $&    $  _{+0.0} ^ {+0.0} $&    $  _{+2.3} ^ {-1.9} $&    +3.0 &    $  _{+0.8} ^ {-0.7} $&    -1.6  \\
&0.090 : 0.160&    $  _{-3.4} ^ {+1.5} $&    $  _{-4.0} ^ {+3.3} $&    $  _{-1.5} ^ {+2.6} $&    $  _{-1.1} ^ {+1.1} $&    +1.0  &+0.5  &$  _{-2.2} ^ {+2.3} $&    $  _{-1.8} ^ {+1.8} $&    $  _{-2.3} ^ {+2.3} $&    $  _{-0.1} ^ {+0.1} $&    $  _{+0.0} ^ {+0.0} $&    $  _{-0.8} ^ {+0.6} $&    +1.8 &    $  _{+1.5} ^ {-1.4} $&    -2.0  \\
&0.160 : 0.320&    $  _{+0.1} ^ {+0.9} $&    $  _{-2.7} ^ {+2.9} $&    $  _{-0.8} ^ {+1.1} $&    $  _{-3.5} ^ {+3.5} $&    +0.6  &+0.3  &$  _{-2.1} ^ {+2.2} $&    $  _{-1.8} ^ {+1.8} $&    $  _{-1.6} ^ {+1.6} $&    $  _{-0.0} ^ {-0.0} $&    $  _{-0.5} ^ {+0.3} $&    $  _{-0.9} ^ {+0.7} $&    -0.5 &    $  _{+1.8} ^ {-1.7} $&    +0.7  \\
&0.320 : 0.700&    $  _{+1.7} ^ {-1.5} $&    $  _{-0.7} ^ {+1.9} $&    $  _{-0.6} ^ {+0.5} $&    $  _{-1.9} ^ {+1.9} $&    +0.2  &-1.5  &$  _{-1.9} ^ {+2.0} $&    $  _{-1.9} ^ {+1.9} $&    $  _{-1.7} ^ {+1.7} $&    $  _{-0.0} ^ {-0.0} $&    $  _{-2.8} ^ {+1.4} $&    $  _{-0.4} ^ {+0.3} $&    -0.6 &    $  _{+1.5} ^ {-1.4} $&    +3.1  \\
\hline
\multirow{5}{*}{9 : 14 } &0.020 : 0.050&    $ _{-6.9} ^ {+11} $&  $ _{+7.3} ^ {-4.8} $&  $ _{-2.5} ^ {+1.5} $&  $ _{+2.5} ^ {-2.5} $&   -0.1 &   -2.1 &  $ _{-2.1} ^ {+2.2} $&    $ _{-2.4} ^ {+2.4} $&    $ _{-6.1} ^ {+6.1} $&    $ _{-0.1} ^ {+0.1} $&    $ _{+0.0} ^ {+0.0} $&    $ _{+4.7} ^ {-4.5} $&     +3.0 &    $ _{+0.7} ^ {-0.7} $&    +1.4  \\
&0.050 : 0.090&    $  _{-3.2} ^ {+3.1} $&    $  _{+1.9} ^ {+0.2} $&    $  _{-1.3} ^ {+1.0} $&    $  _{-2.1} ^ {+2.1} $&    +0.6  &+0.1  &$  _{-2.1} ^ {+2.2} $&    $  _{-2.0} ^ {+2.0} $&    $  _{-2.4} ^ {+2.4} $&    $  _{-0.0} ^ {+0.0} $&    $  _{-0.0} ^ {+0.0} $&    $  _{+1.5} ^ {-1.2} $&    +2.7 &    $  _{+1.2} ^ {-1.2} $&    -1.6  \\
&0.090 : 0.160&    $  _{-0.6} ^ {+0.5} $&    $  _{+0.3} ^ {-0.1} $&    $  _{-0.6} ^ {+1.2} $&    $  _{-1.6} ^ {+1.6} $&    +0.5  &-0.3  &$  _{-2.0} ^ {+2.1} $&    $  _{-1.7} ^ {+1.7} $&    $  _{-1.7} ^ {+1.7} $&    $  _{+0.2} ^ {-0.2} $&    $  _{+0.0} ^ {-0.0} $&    $  _{+0.1} ^ {-0.1} $&    +0.8 &    $  _{+1.7} ^ {-1.6} $&    -2.0  \\
&0.160 : 0.320&    $  _{+0.5} ^ {+0.0} $&    $  _{+0.5} ^ {+0.2} $&    $  _{-0.7} ^ {+0.3} $&    $  _{+0.5} ^ {-0.5} $&    +0.3  &+0.5  &$  _{-2.0} ^ {+2.1} $&    $  _{-1.8} ^ {+1.8} $&    $  _{-1.5} ^ {+1.5} $&    $  _{+0.1} ^ {-0.1} $&    $  _{-0.1} ^ {+0.1} $&    $  _{-0.7} ^ {+0.6} $&    +1.1 &    $  _{+1.9} ^ {-1.8} $&    +0.4  \\
&0.320 : 0.700&    $  _{+3.8} ^ {-2.8} $&    $  _{-0.7} ^ {+1.1} $&    $  _{-0.1} ^ {+0.1} $&    $  _{+0.4} ^ {-0.4} $&    +0.4  &-0.2  &$  _{-1.9} ^ {+1.9} $&    $  _{-1.8} ^ {+1.8} $&    $  _{-1.9} ^ {+1.9} $&    $  _{-0.3} ^ {+0.2} $&    $  _{-1.6} ^ {+0.8} $&    $  _{-0.5} ^ {+0.4} $&    +0.2 &    $  _{+2.0} ^ {-1.8} $&    +2.8  \\
\hline
\multirow{5}{*}{14 : 23 } &0.020 : 0.050&    $ _{-8.8} ^ {+12} $&  $ _{+6.4} ^ {-3.8} $&  $ _{-0.2} ^ {+0.4} $&  $ _{+0.0} ^ {-0.0} $&   +0.3 &   +2.5 &  $ _{-2.0} ^ {+2.1} $&    $ _{-2.4} ^ {+2.4} $&    $ _{-6.0} ^ {+6.0} $&    $ _{-0.2} ^ {+0.2} $&    $ _{+0.0} ^ {+0.0} $&    $ _{+2.4} ^ {-2.7} $&     +4.1 &    $ _{+0.8} ^ {-0.8} $&    +1.7  \\
&0.050 : 0.090&    $  _{-3.9} ^ {+2.5} $&    $  _{+1.5} ^ {-3.8} $&    $  _{-0.3} ^ {+0.3} $&    $  _{+0.9} ^ {-0.9} $&    +0.2  &+1.0  &$  _{-2.0} ^ {+2.1} $&    $  _{-2.0} ^ {+2.0} $&    $  _{-2.3} ^ {+2.3} $&    $  _{-0.1} ^ {+0.1} $&    $  _{+0.0} ^ {+0.0} $&    $  _{+1.1} ^ {-0.9} $&    +2.3 &    $  _{+1.8} ^ {-1.9} $&    -1.7  \\
&0.090 : 0.160&    $  _{+0.4} ^ {+1.3} $&    $  _{+0.8} ^ {-0.4} $&    $  _{-0.3} ^ {+0.2} $&    $  _{+0.3} ^ {-0.3} $&    -1.0  &-1.6  &$  _{-1.9} ^ {+2.0} $&    $  _{-1.7} ^ {+1.7} $&    $  _{-1.6} ^ {+1.6} $&    $  _{+0.1} ^ {-0.1} $&    $  _{-0.0} ^ {+0.0} $&    $  _{-0.4} ^ {+0.3} $&    +0.7 &    $  _{+2.2} ^ {-2.3} $&    -2.0  \\
&0.160 : 0.320&    $  _{+0.5} ^ {-0.9} $&    $  _{-0.2} ^ {-0.6} $&    $  _{-0.0} ^ {+0.0} $&    $  _{-0.6} ^ {+0.6} $&    -0.9  &+1.7  &$  _{-1.9} ^ {+1.9} $&    $  _{-1.7} ^ {+1.7} $&    $  _{-1.4} ^ {+1.4} $&    $  _{-0.2} ^ {+0.1} $&    $  _{-0.3} ^ {+0.2} $&    $  _{-1.3} ^ {+1.0} $&    +0.4 &    $  _{+2.6} ^ {-2.6} $&    +0.4  \\
&0.320 : 0.700&    $  _{+1.3} ^ {-2.9} $&    $  _{-1.7} ^ {-0.3} $&    $  _{-0.0} ^ {+0.0} $&    $  _{-0.1} ^ {+0.1} $&    -0.3  &-3.1  &$  _{-1.8} ^ {+1.9} $&    $  _{-1.8} ^ {+1.8} $&    $  _{-1.8} ^ {+1.8} $&    $  _{-0.3} ^ {+0.2} $&    $  _{-1.9} ^ {+1.0} $&    $  _{-0.4} ^ {+0.3} $&    -1.5 &    $  _{+2.7} ^ {-2.7} $&    +2.9  \\
\hline
\multirow{5}{*}{23 : 45 } &0.020 : 0.050&    $ _{-8.4} ^ {+9.5} $&  $ _{+12} ^ {-6.7} $&  $ _{-0.0} ^ {+0.0} $&  $ _{+0.0} ^ {-0.0} $&   +1.3 &   -13 &  $ _{-1.8} ^ {+1.9} $&    $ _{-2.2} ^ {+2.2} $&    $ _{-7.0} ^ {+7.0} $&    $ _{+0.3} ^ {-0.3} $&    $ _{+0.0} ^ {+0.0} $&    $ _{+2.4} ^ {-2.3} $&     +3.7 &    $ _{+0.6} ^ {-0.6} $&    +1.7  \\
&0.050 : 0.090&    $  _{-4.6} ^ {+4.3} $&    $  _{+4.3} ^ {-4.8} $&    $  _{-0.0} ^ {+0.0} $&    $  _{+0.0} ^ {-0.0} $&    -0.3  &-0.2  &$  _{-1.7} ^ {+1.8} $&    $  _{-1.9} ^ {+1.9} $&    $  _{-2.3} ^ {+2.3} $&    $  _{-0.0} ^ {+0.0} $&    $  _{+0.0} ^ {+0.0} $&    $  _{+0.5} ^ {-0.4} $&    +0.1 &    $  _{+1.6} ^ {-1.5} $&    -1.4  \\
&0.090 : 0.160&    $  _{-1.1} ^ {+0.9} $&    $  _{+1.8} ^ {-1.4} $&    $  _{-0.0} ^ {+0.0} $&    $  _{+0.0} ^ {-0.0} $&    +0.3  &-0.1  &$  _{-1.7} ^ {+1.8} $&    $  _{-1.6} ^ {+1.6} $&    $  _{-1.5} ^ {+1.5} $&    $  _{-0.3} ^ {+0.2} $&    $  _{+0.0} ^ {-0.0} $&    $  _{+0.3} ^ {-0.3} $&    +0.0 &    $  _{+1.7} ^ {-1.7} $&    -2.1  \\
&0.160 : 0.320&    $  _{+1.2} ^ {-1.0} $&    $  _{-0.4} ^ {-0.3} $&    $  _{-0.0} ^ {+0.0} $&    $  _{+0.0} ^ {-0.0} $&    +0.2  &+0.1  &$  _{-1.7} ^ {+1.8} $&    $  _{-1.6} ^ {+1.6} $&    $  _{-1.3} ^ {+1.3} $&    $  _{-0.2} ^ {+0.1} $&    $  _{-0.2} ^ {+0.1} $&    $  _{-0.1} ^ {+0.0} $&    +0.7 &    $  _{+1.8} ^ {-1.8} $&    +0.4  \\
&0.320 : 0.700&    $  _{+3.5} ^ {-2.6} $&    $  _{+1.1} ^ {+0.0} $&    $  _{-0.0} ^ {+0.0} $&    $  _{+0.0} ^ {-0.0} $&    +1.0  &+0.0  &$  _{-1.7} ^ {+1.8} $&    $  _{-1.8} ^ {+1.8} $&    $  _{-1.5} ^ {+1.5} $&    $  _{+0.1} ^ {-0.2} $&    $  _{-1.3} ^ {+0.6} $&    $  _{-0.2} ^ {+0.2} $&    +2.1 &    $  _{+1.7} ^ {-1.7} $&    +2.4  \\
\hline
\multirow{5}{*}{45 : 100 } &0.020 : 0.050&    $ _{-6.4} ^ {+26} $&  $ _{+18} ^ {-9.7} $&  $ _{-0.0} ^ {+0.0} $&  $ _{+0.0} ^ {-0.0} $&   -0.1 &   +2.3 &  $ _{-1.4} ^ {+1.5} $&    $ _{-2.0} ^ {+2.0} $&    $ _{-13} ^ {+13} $&    $ _{+0.0} ^ {+0.0} $&    $ _{+0.0} ^ {+0.0} $&    $ _{+3.4} ^ {-2.6} $&     +5.0 &    $ _{+2.0} ^ {-2.2} $&    +3.5  \\
&0.050 : 0.090&    $  _{-3.5} ^ {+4.2} $&    $  _{+3.7} ^ {-4.0} $&    $  _{-0.0} ^ {+0.0} $&    $  _{+0.0} ^ {-0.0} $&    -2.0  &-0.1  &$  _{-1.5} ^ {+1.5} $&    $  _{-1.8} ^ {+1.8} $&    $  _{-3.1} ^ {+3.1} $&    $  _{-0.1} ^ {+0.1} $&    $  _{+0.0} ^ {+0.0} $&    $  _{+0.0} ^ {-0.2} $&    +3.3 &    $  _{+1.3} ^ {-1.4} $&    -1.6  \\
&0.090 : 0.160&    $  _{-0.7} ^ {+1.1} $&    $  _{+2.4} ^ {-2.0} $&    $  _{-0.0} ^ {+0.0} $&    $  _{+0.0} ^ {-0.0} $&    -0.8  &-0.3  &$  _{-1.4} ^ {+1.5} $&    $  _{-1.5} ^ {+1.5} $&    $  _{-1.7} ^ {+1.7} $&    $  _{-0.4} ^ {+0.3} $&    $  _{-0.0} ^ {+0.0} $&    $  _{+0.0} ^ {-0.0} $&    +1.5 &    $  _{+1.7} ^ {-1.9} $&    -2.1  \\
&0.160 : 0.320&    $  _{-0.1} ^ {-0.7} $&    $  _{+1.9} ^ {-1.5} $&    $  _{-0.0} ^ {+0.0} $&    $  _{+0.0} ^ {-0.0} $&    +0.3  &+0.5  &$  _{-1.5} ^ {+1.6} $&    $  _{-1.5} ^ {+1.5} $&    $  _{-1.4} ^ {+1.4} $&    $  _{-0.7} ^ {+0.6} $&    $  _{-0.0} ^ {+0.0} $&    $  _{+0.0} ^ {-0.0} $&    +3.0 &    $  _{+2.1} ^ {-2.3} $&    +0.2  \\
&0.320 : 0.700&    $  _{+3.2} ^ {-2.8} $&    $  _{-0.3} ^ {-0.8} $&    $  _{-0.0} ^ {+0.0} $&    $  _{+0.0} ^ {-0.0} $&    -0.6  &-2.4  &$  _{-1.6} ^ {+1.7} $&    $  _{-1.6} ^ {+1.6} $&    $  _{-1.6} ^ {+1.6} $&    $  _{-0.7} ^ {+0.5} $&    $  _{-0.9} ^ {+0.4} $&    $  _{+0.7} ^ {-0.6} $&    +2.4 &    $  _{+2.5} ^ {-2.7} $&    +2.9  \\
\hline
\multirow{3}{*}{100 : 158 } &0.020 : 0.350&    $ _{-0.6} ^ {+1.7} $&  $ _{+4.4} ^ {-1.3} $&  $ _{-0.0} ^ {+0.0} $&  $ _{+0.0} ^ {-0.0} $&   -2.2 &   +4.4 &  $ _{-1.3} ^ {+1.3} $&    $ _{-1.4} ^ {+1.4} $&    $ _{-1.8} ^ {+1.8} $&    $ _{-1.1} ^ {+0.9} $&    $ _{-0.0} ^ {+0.0} $&    $ _{+0.2} ^ {-0.2} $&     +1.3 &    $ _{+1.2} ^ {-1.3} $&    -0.2  \\
&0.350 : 0.700&    $  _{+0.8} ^ {-5.3} $&    $  _{-0.9} ^ {-2.5} $&    $  _{-0.0} ^ {+0.0} $&    $  _{+0.0} ^ {-0.0} $&    +1.8  &+4.4  &$  _{-1.4} ^ {+1.4} $&    $  _{-1.5} ^ {+1.5} $&    $  _{-2.9} ^ {+2.9} $&    $  _{-3.2} ^ {+2.6} $&    $  _{-1.1} ^ {+0.5} $&    $  _{-0.4} ^ {+0.1} $&    -0.2 &    $  _{+1.1} ^ {-1.3} $&    +2.9  \\
\hline
\multirow{3}{*}{158 : 251 } &0.020 : 0.300&    $ _{-0.8} ^ {+0.3} $&  $ _{+2.0} ^ {-4.3} $&  $ _{-0.0} ^ {+0.0} $&  $ _{+0.0} ^ {-0.0} $&   -1.9 &   -0.3 &  $ _{-1.2} ^ {+1.2} $&    $ _{-1.3} ^ {+1.3} $&    $ _{-3.0} ^ {+3.0} $&    $ _{-1.0} ^ {+0.8} $&    $ _{-0.1} ^ {+0.1} $&    $ _{-0.6} ^ {+0.4} $&     -1.3 &    $ _{+0.8} ^ {-0.9} $&    +0.1  \\
&0.300 : 0.700&    $  _{+2.6} ^ {-2.6} $&    $  _{+0.6} ^ {-0.2} $&    $  _{-0.0} ^ {+0.0} $&    $  _{+0.0} ^ {-0.0} $&    +1.8  &+1.7  &$  _{-1.3} ^ {+1.4} $&    $  _{-1.4} ^ {+1.4} $&    $  _{-3.0} ^ {+3.0} $&    $  _{-0.7} ^ {+0.1} $&    $  _{-1.0} ^ {+0.5} $&    $  _{-0.5} ^ {+0.3} $&    +1.1 &    $  _{+0.9} ^ {-1.0} $&    +3.0  \\
\hline
\multirow{3}{*}{251 : 1000 } &0.020 : 0.275&    $ _{-2.2} ^ {-0.6} $&  $ _{+5.9} ^ {-5.8} $&  $ _{-0.0} ^ {+0.0} $&  $ _{+0.0} ^ {-0.0} $&   -4.6 &   -0.0 &  $ _{-1.2} ^ {+1.2} $&    $ _{-1.3} ^ {+1.3} $&    $ _{-4.6} ^ {+4.6} $&    $ _{-1.3} ^ {+0.8} $&    $ _{+0.0} ^ {+0.0} $&    $ _{+0.9} ^ {-1.0} $&     -3.3 &    $ _{-0.1} ^ {+0.1} $&    -0.3  \\
&0.275 : 0.700&    $  _{+4.3} ^ {-0.8} $&    $  _{+3.8} ^ {-1.7} $&    $  _{-0.0} ^ {+0.0} $&    $  _{+0.0} ^ {-0.0} $&    +2.5  &+3.5  &$  _{-1.3} ^ {+1.4} $&    $  _{-1.4} ^ {+1.4} $&    $  _{-3.6} ^ {+3.6} $&    $  _{-0.1} ^ {-0.1} $&    $  _{-1.4} ^ {+0.7} $&    $  _{+0.4} ^ {-0.5} $&    -1.4 &    $  _{+0.9} ^ {-1.0} $&    +0.9  \\
\hline

\end{tabular}
\caption{ Individual systematical uncertainties as defined in Section~\ref{sub:syst} for the double-differential cross sections 
in bins of  ~$Q^2$ and $y$. The uncertainty $\delta_8$ and $\delta_{10}$ are not reported as $\delta_8$ is constant ($+2\,\%$) 
and $\delta_{10}$ was found to be negligible. 
The overall normalisation uncertanties $\delta_{18}=\pm1.9 \%$ and $\delta_{19}=\pm 1.5\%$ are also not listed. }
\label{t:syst}
\end{center}
\end{table}

\begin{table}
\begin{center}
{\footnotesize
\tabcolsep2mm
\begin{tabular}{|c|c|r|r|r|r|r|r|} \hline
$Q^2$     & $x$ & $\delta_{m_c}$ & $\delta_{\mu}$ & $\delta_{\alpha_s}$  & $\delta_{\alpha_K}$ & $\delta_{k_T}$  & $\delta_{\rm b}$ \\
(\gev$^{2}$) &     & (\%)           & (\%)           & (\%)                 & (\%)                & (\%)              & (\%) \\ \hline
      &  0.00160 & $_{ -5.6}^{ +8.3}$ & $_{+14}^{ -6.5}$ & $_{ +0.2}^{ +0.6}$ & $_{ +7.5}^{ -4.5}$& $_{ +0.6}^{ -0.2}$ & $\pm  0.0$ \\ 
      &  0.00080 & $_{ +1.0}^{ +0.3}$ & $_{ +7.8}^{ -3.9}$ & $_{ +0.6}^{ -0.3}$ & $_{ +6.1}^{ -3.5}$& $_{ +1.3}^{ -1.3}$ & $\pm  0.3$ \\ 
     7 &  0.00050 & $_{ +2.0}^{ -1.3}$ & $_{ +5.0}^{ -3.2}$ & $_{ +0.8}^{ +0.1}$ & $_{ +6.5}^{ -2.9}$& $_{ +1.8}^{ -1.3}$ & $\pm  0.6$ \\ 
      &  0.00030 & $_{ +3.3}^{ -3.2}$ & $_{ +0.2}^{ -1.4}$ & $_{ +1.2}^{ -0.9}$ & $_{ +5.9}^{ -2.6}$& $_{ +2.2}^{ -2.5}$ & $\pm  1.0$ \\ 
      &  0.00013 & $_{ +5.7}^{ -3.7}$ & $_{ -6.3}^{ +4.7}$ & $_{ +2.5}^{ -1.6}$ & $_{ +5.9}^{ -2.4}$& $_{ +4.2}^{ -4.0}$ & $\pm  2.1$ \\ \hline 
     &  0.00300 & $_{ -6.2}^{ +9.5}$ & $_{+15}^{ -6.5}$ & $_{ +0.0}^{ +1.4}$ & $_{ +8.0}^{ -3.6}$& $_{ +0.1}^{ +1.6}$ & $\pm  0.0$ \\ 
     &  0.00150 & $_{ -1.1}^{ +0.1}$ & $_{ +7.8}^{ -5.4}$ & $_{ -0.6}^{ -0.1}$ & $_{ +5.3}^{ -3.3}$& $_{ -0.1}^{ -0.7}$ & $\pm  0.3$ \\ 
    12 &  0.00080 & $_{ +0.8}^{ -0.6}$ & $_{ +5.7}^{ -3.8}$ & $_{ +0.1}^{ -0.1}$ & $_{ +5.5}^{ -2.5}$& $_{ +1.0}^{ -0.9}$ & $\pm  0.7$ \\ 
     &  0.00050 & $_{ +2.2}^{ -2.5}$ & $_{ +2.0}^{ -2.5}$ & $_{ +0.0}^{ -0.5}$ & $_{ +5.0}^{ -2.4}$& $_{ +1.2}^{ -1.9}$ & $\pm  1.2$ \\ 
     &  0.00022 & $_{ +3.9}^{ -3.2}$ & $_{ -4.4}^{ +3.1}$ & $_{ +1.7}^{ -1.6}$ & $_{ +5.4}^{ -2.2}$& $_{ +1.8}^{ -3.1}$ & $\pm  2.1$ \\ \hline 
     &  0.00450 & $_{ -6.1}^{ +8.8}$ & $_{+13}^{ -6.5}$ & $_{ +0.7}^{ +0.9}$ & $_{ +6.2}^{ -3.2}$& $_{ -1.0}^{ +1.1}$ & $\pm  0.1$ \\ 
     &  0.00250 & $_{ -0.9}^{ +0.3}$ & $_{ +7.0}^{ -5.7}$ & $_{ -0.6}^{ +0.4}$ & $_{ +3.7}^{ -3.2}$& $_{ -0.6}^{ -0.3}$ & $\pm  0.4$ \\ 
    18 &  0.00135 & $_{ +0.8}^{ -0.4}$ & $_{ +6.1}^{ -4.4}$ & $_{ +0.1}^{ +0.6}$ & $_{ +4.8}^{ -2.4}$& $_{ +0.6}^{ -0.5}$ & $\pm  0.8$ \\ 
     &  0.00080 & $_{ +1.0}^{ -1.5}$ & $_{ +3.2}^{ -4.0}$ & $_{ +0.3}^{ +0.3}$ & $_{ +4.4}^{ -1.9}$& $_{ +0.7}^{ -0.9}$ & $\pm  1.7$ \\ 
     &  0.00035 & $_{ +2.7}^{ -3.0}$ & $_{ -3.7}^{ +1.8}$ & $_{ +1.0}^{ -1.0}$ & $_{ +4.5}^{ -2.5}$& $_{ +1.4}^{ -2.9}$ & $\pm  2.9$ \\ \hline 
     &  0.00800 & $_{ -7.3}^{ +8.4}$ & $_{+11}^{ -7.0}$ & $_{ -0.5}^{ +0.6}$ & $_{ +5.1}^{ -3.5}$& $_{ -1.7}^{ +0.3}$ & $\pm  0.1$ \\ 
     &  0.00550 & $_{ -0.0}^{ +1.3}$ & $_{ +8.4}^{ -5.8}$ & $_{ -0.3}^{ +0.5}$ & $_{ +3.2}^{ -1.9}$& $_{ -0.3}^{ +0.3}$ & $\pm  0.3$ \\ 
    32 &  0.00240 & $_{ +0.5}^{ +0.5}$ & $_{ +6.4}^{ -3.6}$ & $_{ +0.3}^{ -0.1}$ & $_{ +3.9}^{ -1.7}$& $_{ +0.0}^{ -0.2}$ & $\pm  0.9$ \\ 
     &  0.00140 & $_{ +1.3}^{ -0.5}$ & $_{ +4.6}^{ -3.5}$ & $_{ +0.1}^{ +0.2}$ & $_{ +3.9}^{ -1.6}$& $_{ +0.6}^{ -0.4}$ & $\pm  1.9$ \\ 
     &  0.00080 & $_{ +3.0}^{ -2.9}$ & $_{ -1.6}^{ -0.4}$ & $_{ +0.5}^{ -0.8}$ & $_{ +3.6}^{ -2.2}$& $_{ +1.1}^{ -2.2}$ & $\pm  2.7$ \\ \hline 
     &  0.01500 & $_{ -6.5}^{ +9.3}$ & $_{+10}^{ -5.2}$ & $_{ +0.4}^{ +0.6}$ & $_{ +6.2}^{ -1.8}$& $_{ +0.4}^{ +1.6}$ & $\pm  0.0$ \\ 
     &  0.00800 & $_{ -1.7}^{ +0.6}$ & $_{ +6.0}^{ -4.8}$ & $_{ -0.7}^{ -0.3}$ & $_{ +2.3}^{ -1.9}$& $_{ -0.6}^{ -0.1}$ & $\pm  0.9$ \\ 
    60 &  0.00500 & $_{ +0.8}^{ -0.2}$ & $_{ +5.2}^{ -3.9}$ & $_{ -0.0}^{ +0.1}$ & $_{ +2.7}^{ -1.4}$& $_{ +0.3}^{ -0.3}$ & $\pm  2.3$ \\ 
     &  0.00320 & $_{ +1.4}^{ -0.9}$ & $_{ +5.0}^{ -3.7}$ & $_{ -0.2}^{ -0.1}$ & $_{ +2.8}^{ -1.6}$& $_{ +0.0}^{ -0.4}$ & $\pm  2.7$ \\ 
     &  0.00140 & $_{ +1.8}^{ -2.4}$ & $_{ +1.3}^{ -1.5}$ & $_{ -0.0}^{ -0.1}$ & $_{ +2.8}^{ -1.8}$& $_{ +0.6}^{ -1.3}$ & $\pm  4.4$ \\ \hline 
       &  0.01000 & $_{ +0.8}^{ +0.2}$ & $_{ +5.3}^{ -4.6}$ & $_{ +0.1}^{ +0.4}$ & $_{ +2.3}^{ -1.5}$& $_{ +0.3}^{ +0.0}$ & $\pm  3.9$ \\ 
   120 &  0.00200 & $_{ +1.3}^{ -0.8}$ & $_{ +2.3}^{ -2.0}$ & $_{ -0.5}^{ +0.4}$ & $_{ +1.9}^{ -1.3}$& $_{ +0.8}^{ -1.0}$ & $\pm  6.3$ \\ \hline 
       &  0.01300 & $_{ -0.1}^{ -0.1}$ & $_{ +3.8}^{ -3.7}$ & $_{ -0.1}^{ +0.4}$ & $_{ +1.4}^{ -0.9}$& $_{ +0.0}^{ +0.1}$ & $\pm  3.8$ \\ 
   200 &  0.00500 & $_{ +1.3}^{ -1.9}$ & $_{ +3.8}^{ -3.8}$ & $_{ -0.6}^{ -0.3}$ & $_{ +1.2}^{ -1.5}$& $_{ +0.1}^{ -0.1}$ & $\pm  6.5$ \\ \hline 
       &  0.02500 & $_{ +0.4}^{ -0.5}$ & $_{ +3.4}^{ -3.8}$ & $_{ -0.0}^{ -0.4}$ & $_{ +1.2}^{ -0.7}$& $_{ -0.4}^{ +0.4}$ & $\pm  4.6$ \\ 
   350 &  0.01000 & $_{ +1.3}^{ -0.2}$ & $_{ +3.7}^{ -2.8}$ & $_{ +0.3}^{ +0.0}$ & $_{ +0.9}^{ -0.6}$& $_{ +0.1}^{ +0.0}$ & $\pm  8.7$ \\ \hline 
\end{tabular}
}
\end{center}
\caption{Breakdown of the theoretical uncertainty on $\sigma_{\rm red}^{c\bar{c}}(x,Q^2)$, showing the uncertainty from the variation of charm mass ($\delta_{m_c}$), of the renormalisation and factorisation scales  ($\delta_{\mu}$), of $\alpha_S$ ($\delta_{\alpha_s}$),  of the fragmentation function ($\delta_{\alpha_K}$), of the transverse fragmentation ($\delta_{k_T}$),  and of the expected  beauty component ($\delta_{\rm b}$). The upper (lower) value gives the effect of a positive (negative) variation of the parameter.}
\label{t:theo}
\end{table}

\begin{table}
\begin{center}
\tabcolsep1.9mm
\begin{tabular}{|c|c|c|r|r|r|r|}
\hline
$Q^2$ & $x$ & $\sigma_{\rm red}^{c\bar{c}}$ & $\delta_{\rm stat.}$ &
$\delta_{\rm syst.}$ & $\delta_{\rm theo.}$ & ${\cal A}_{ps}$ \\ 
(\gev$^{2}$) &   &   & (\%) & (\%) & (\%) & (\%)\\
\hline
        &  0.00160 &    0.057 &    23 & $_{- 20}  ^{+ 19}$ & $_{-  9.7}^{+   18}$ & 0.248 \\ 
        &  0.00080 &    0.124 &    10 & $_{- 11}  ^{+ 11}$ & $_{-  5.4}^{+   10}$ & 0.412 \\ 
      7 &  0.00050 &    0.166 &   6.1 & $_{-  7.1}^{+  6.8}$ & $_{-  4.7}^{+  8.7}$ & 0.480 \\ 
        &  0.00030 &    0.191 &   5.4 & $_{-  6.0}^{+  6.7}$ & $_{-  5.2}^{+  7.3}$ & 0.481 \\ 
        &  0.00013 &    0.258 &   7.1 & $_{-  5.7}^{+  6.6}$ & $_{-  9.0}^{+   11}$ & 0.327 \\ \hline

        &  0.00300 &    0.098 &    14 & $_{-   12}^{+   17}$ & $_{-  9.7}^{+   19}$ & 0.280 \\ 
        &  0.00150 &    0.153 &   6.6 & $_{-  6.0}^{+  7.1}$ & $_{-  6.5}^{+  9.4}$ & 0.462 \\ 
     12 &  0.00080 &    0.177 &   5.9 & $_{-  4.6}^{+  4.7}$ & $_{-  4.7}^{+  8.1}$ & 0.536 \\ 
        &  0.00050 &    0.244 &   5.2 & $_{-  3.8}^{+  4.6}$ & $_{-  4.9}^{+  6.0}$ & 0.538 \\ 
        &  0.00022 &    0.350 &   7.5 & $_{-  5.2}^{+  6.9}$ & $_{-  7.1}^{+  8.1}$ & 0.363 \\ \hline

        &  0.00450 &    0.081 &    15 & $_{-   12}^{+   16}$ & $_{-  9.5}^{+   17}$ & 0.286 \\ 
        &  0.00250 &    0.169 &   6.5 & $_{-  7.2}^{+  6.2}$ & $_{-  6.7}^{+  8.0}$ & 0.499 \\ 
     18 &  0.00135 &    0.202 &   5.7 & $_{-  4.8}^{+  4.7}$ & $_{-  5.1}^{+  7.9}$ & 0.578 \\ 
        &  0.00080 &    0.224 &   6.1 & $_{-  4.6}^{+  5.1}$ & $_{-  5.0}^{+  5.9}$ & 0.595 \\ 
        &  0.00035 &    0.343 &   7.8 & $_{-  7.1}^{+  6.1}$ & $_{-  6.8}^{+  6.5}$ & 0.404 \\ \hline

        &  0.00800 &    0.068 &    29 & $_{-   18}^{+   18}$ & $_{-   11}^{+   15}$ & 0.258 \\ 
        &  0.00550 &    0.160 &   7.0 & $_{-  7.9}^{+  7.5}$ & $_{-  6.2}^{+  9.1}$ & 0.523 \\ 
     32 &  0.00240 &    0.238 &   5.5 & $_{-  4.4}^{+  4.5}$ & $_{-  4.1}^{+  7.6}$ & 0.613 \\ 
        &  0.00140 &    0.277 &   5.3 & $_{-  3.5}^{+  4.3}$ & $_{-  4.4}^{+  6.5}$ & 0.649 \\ 
        &  0.00080 &    0.412 &   6.4 & $_{-  4.7}^{+  6.8}$ & $_{-  5.4}^{+  5.6}$ & 0.470 \\ \hline

        &  0.01500 &    0.068 &    38 & $_{-   18}^{+   35}$ & $_{-  8.6}^{+   15}$ & 0.182 \\ 
        &  0.00800 &    0.176 &   9.7 & $_{-  7.3}^{+  8.1}$ & $_{-  5.6}^{+  6.6}$ & 0.508 \\ 
     60 &  0.00500 &    0.169 &   8.8 & $_{-  4.9}^{+  5.1}$ & $_{-  4.7}^{+  6.4}$ & 0.624 \\ 
        &  0.00320 &    0.273 &   6.0 & $_{-  4.1}^{+  5.6}$ & $_{-  5.0}^{+  6.5}$ & 0.682 \\ 
        &  0.00140 &    0.359 &   8.2 & $_{-  6.1}^{+  7.2}$ & $_{-  5.7}^{+  5.7}$ & 0.564 \\ \hline

        &  0.01000 &    0.141 &    12 & $_{-  4.5}^{+  8.2}$ & $_{-  6.2}^{+  7.0}$ & 0.536 \\ 
    120 &  0.00200 &    0.329 &    18 & $_{-  8.8}^{+  8.5}$ & $_{-  6.9}^{+  7.1}$ & 0.638 \\ \hline

        &  0.01300 &    0.191 &    16 & $_{-  6.8}^{+  5.1}$ & $_{-  5.4}^{+  5.6}$ & 0.508 \\ 
    200 &  0.00500 &    0.275 &    19 & $_{-  5.5}^{+  7.4}$ & $_{-  8.0}^{+  7.8}$ & 0.682 \\ \hline

        &  0.02500 &    0.113 &    27 & $_{-   11}^{+  8.9}$ & $_{-  6.0}^{+  5.8}$ & 0.474 \\ 
    350 &  0.01000 &    0.234 &    24 & $_{-  6.0}^{+   10}$ & $_{-  9.2}^{+  9.6}$ & 0.696 \\ \hline
\hline
\end{tabular}
\end{center}
\caption{The reduced cross-section $\sigma_{\rm red}^{c\bar{c}}(x,Q^2)$ with statistical, systematic and theoretical uncertainties. The last column shows the kinematical acceptance.}
\label{t:sig_red}
\end{table}

%% file: DESY-13-054-fig.tex

%
\begin{figure}[p]
	\vfill
	\begin{center}
	\includegraphics[scale=0.85]{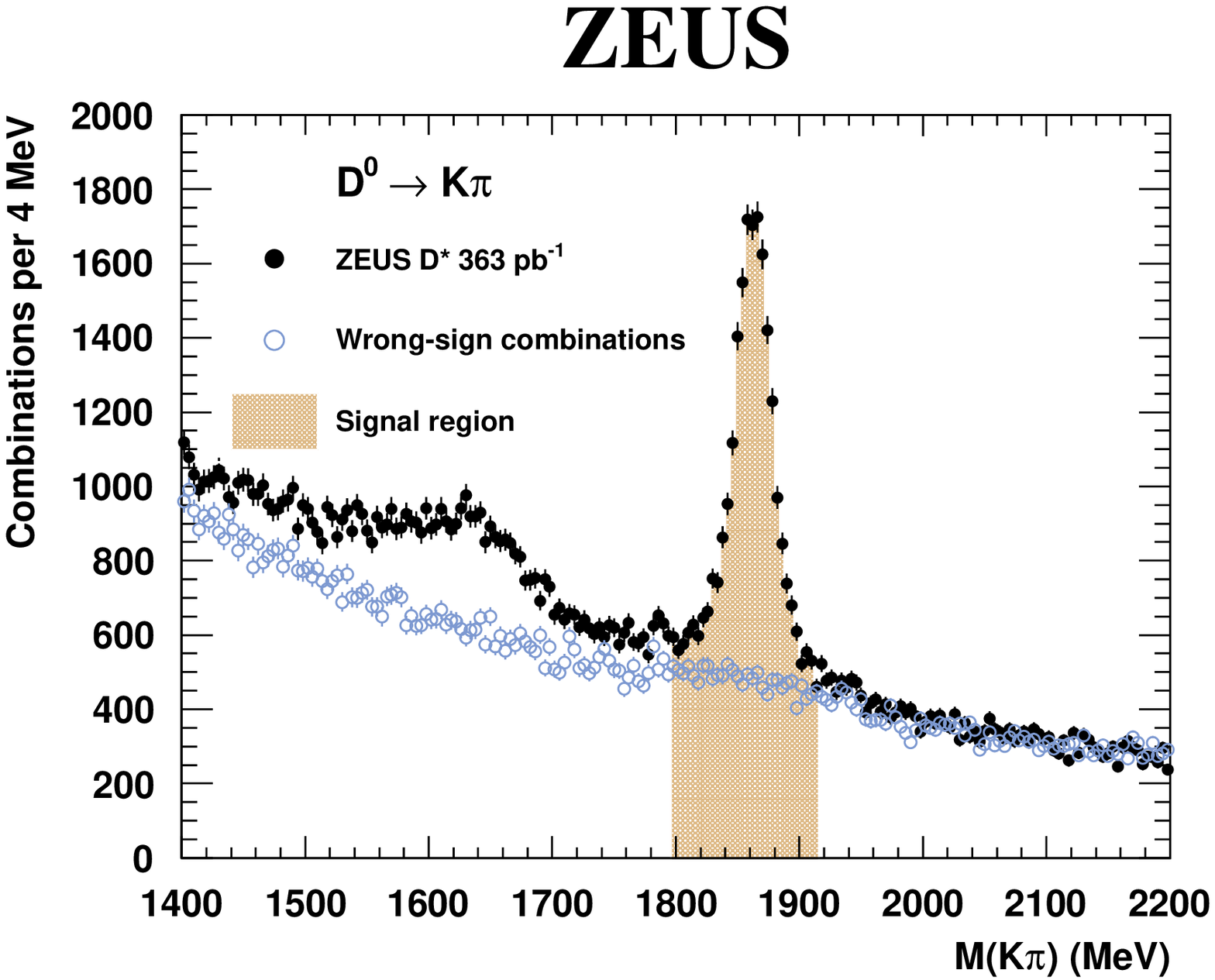}
	\end{center}
	\caption
	{
	\textit{Distribution of  $M (K\pi)$ for $D^{*\pm}$
          candidates with $143.2\!<\! \Delta M \!< \!147.7$~MeV (filled circles)
	  and for wrong-sign combinations (empty circles). The $D^{0}$
          signal region is marked as a shaded area.}
	}
	\label{f:dzero}
	\vfill
\end{figure}

%
\begin{figure}[p]
	\vfill
	\begin{center}
	\includegraphics[scale=0.9]{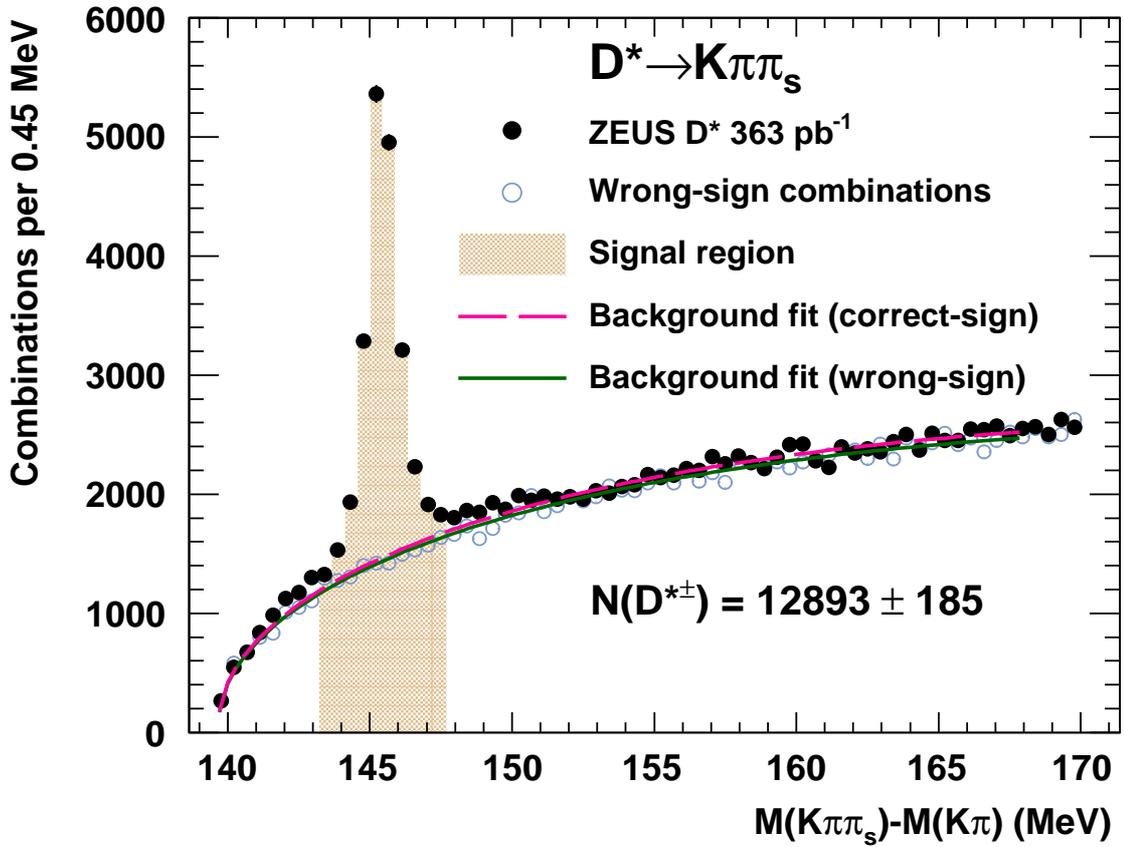}
        \end{center}
	\caption
	{
	\textit{Distribution of the mass difference, $\Delta
          M=M(K\pi\pi_{s})-M(K\pi)$, for the $D^{*\pm}$ candidates 
          with $1.80<M(K\pi)<1.92$~GeV (filled circles)
	and for wrong-sign combinations (empty circles). The background fit
        described in the text is shown as a dashed  (continuous) line for
        correct-sign (wrong-sign) combinations. The $D^{*\pm}$ signal region is marked as a shaded area.}
	}
	\label{f:dstar}
	\vfill
\end{figure}
%
\begin{figure}[p]
	\vfill
	\begin{center}
	\includegraphics[width=\textwidth]{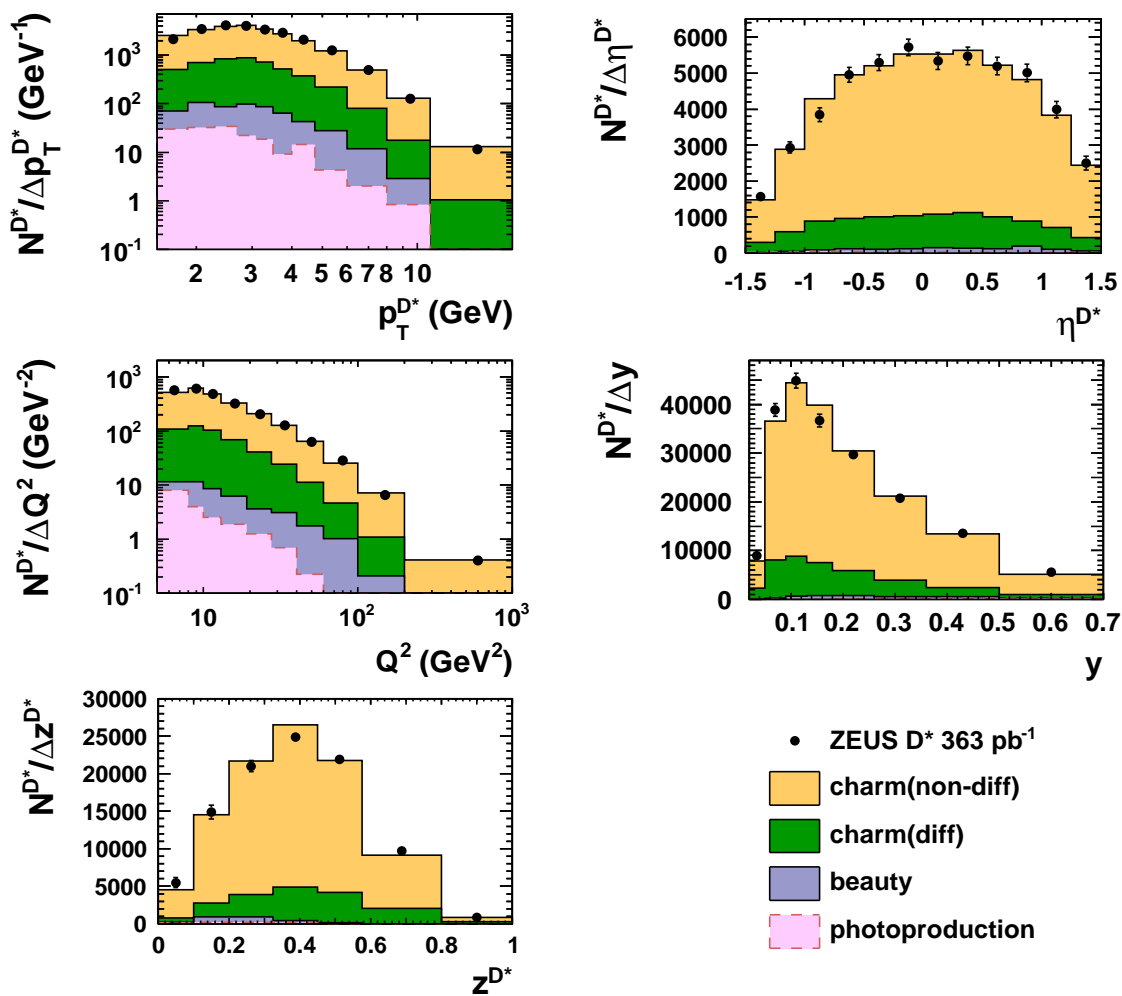}
        \end{center}
	\caption
	{\textit{Number of reconstructed $D^{*\pm}$ (filled circles), divided by
            bin size, as a  function of  ~$p_T^{D^*}$, $\eta^{D^*}$, $Q^2$, $y$ and $z^{D^*}$. 
	Data are compared to a MC  mixture containing non-diffractive and diffractive charm
        production in DIS, beauty production, and  charm
        photoproduction.  The sum of the MC
        samples is normalised to the number of $D^{*\pm}$ in the data.}
	}
	\label{f:control}
	\vfill
\end{figure}

%
\begin{figure}[p]
    \vfill
    \begin{center}
    \includegraphics[width=\textwidth]{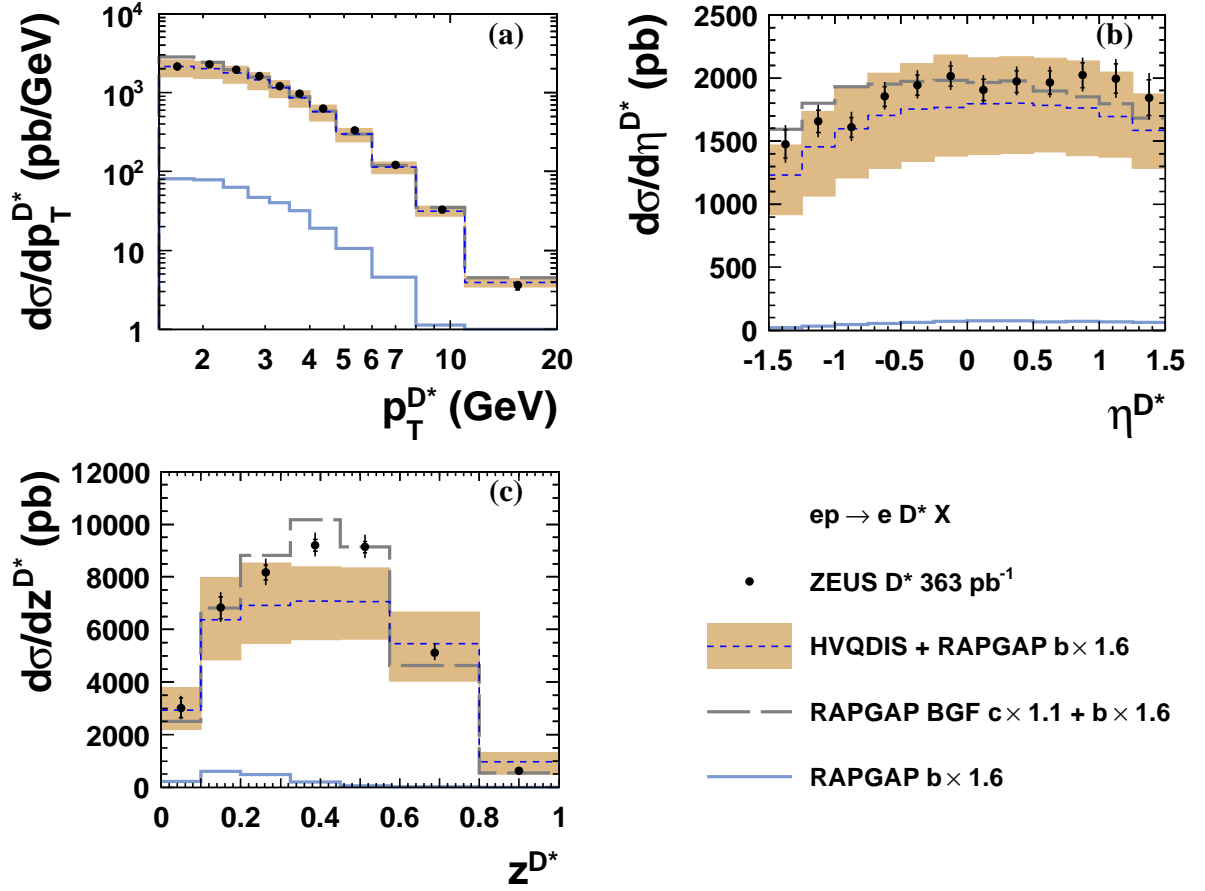}
    \end{center}
    \caption
    {
    \textit{Differential $D^{*\pm} $ cross sections as a function of (a) $p_{T}^{D^{*}}$, (b) 
$\eta^{D^{*}}$ and 
      (c) $z^{D^{*}}$ (filled circles).  The error bars show the statistical and 
      systematic uncertainties added in quadrature, the inner bars show the
      statistical uncertainties alone. Also shown are NLO QCD predictions calculated 
      using {\sc Hvqdis} (dashed line and shaded area for the uncertainties) and 
      {\sc Rapgap} MC prediction for charm creation via
      boson-gluon fusion (long-dashed line). The contribution from b-quark
      decays, calculated with the {\sc Rapgap} MC (continuous line), is
      included in the predictions. The MC cross sections for charm (beauty)
      are scaled by $1.1$ ($1.6$) as described in the text. }
  }
    \label{f:single_2}
    \vfill
\end{figure}
%
\begin{figure}[p]
	\vfill
	\begin{center}
        \includegraphics[width=\textwidth]{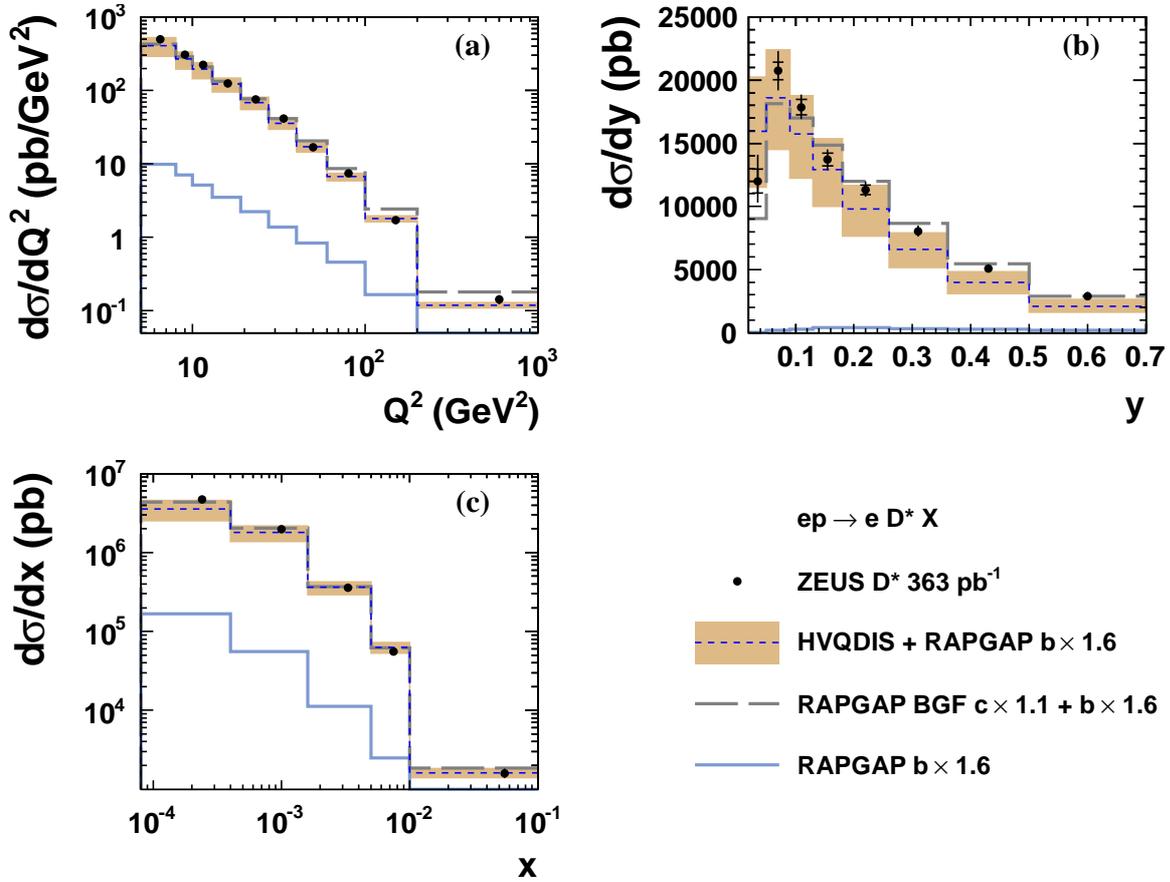}
	\end{center}
	\caption
	{
          \textit{Differential $D^{*\pm} $ cross sections as a function of (a) $Q^{2}$, (b) $y$ and (c) 
$x$. Other details as in Fig.~\protect\ref{f:single_2}.}
        }
	\label{f:single_1}
	\vfill
\end{figure}

%
\begin{figure}[p]
	\vfill
 	\begin{center}
 	\includegraphics[width=\textwidth]{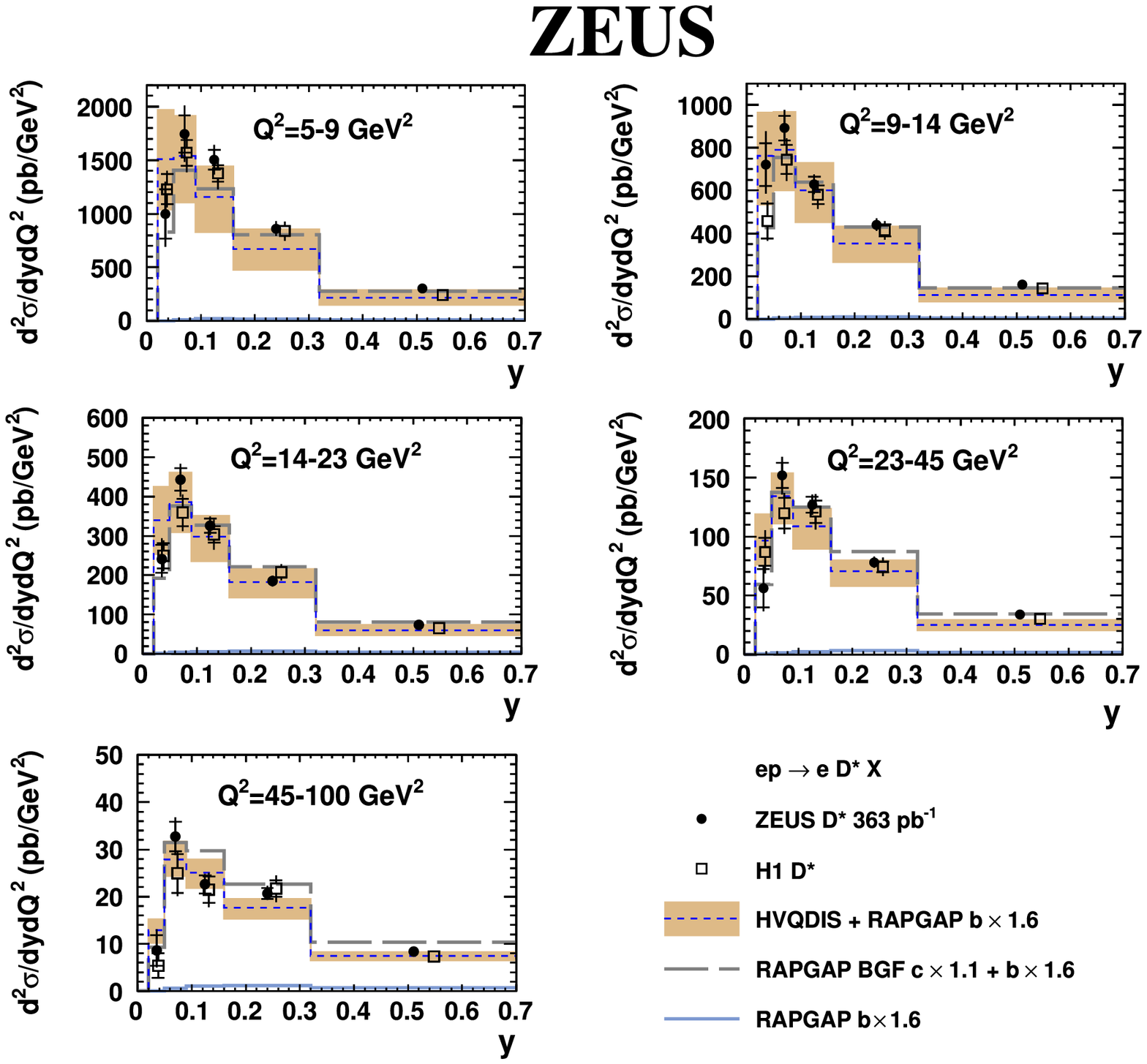}
        \end{center}
	\caption
	{
	\textit{Double-differential $D^{*\pm} $ cross sections as a function of $Q^{2}$ and $y$ 
          for $5<Q^{2}<100$ GeV\,$^{\,2}$ (filled circles).  The
          measurements from the H1 collaboration (empty squares) are
          also shown~\protect\cite{Aaron:2011gp}. Other details as in Fig.~\protect\ref{f:single_2}.}
	}
	\label{f:double_1}
	\vfill
\end{figure}
%
\begin{figure}[p]
    \vfill
    \begin{center}
    \includegraphics[width=\textwidth]{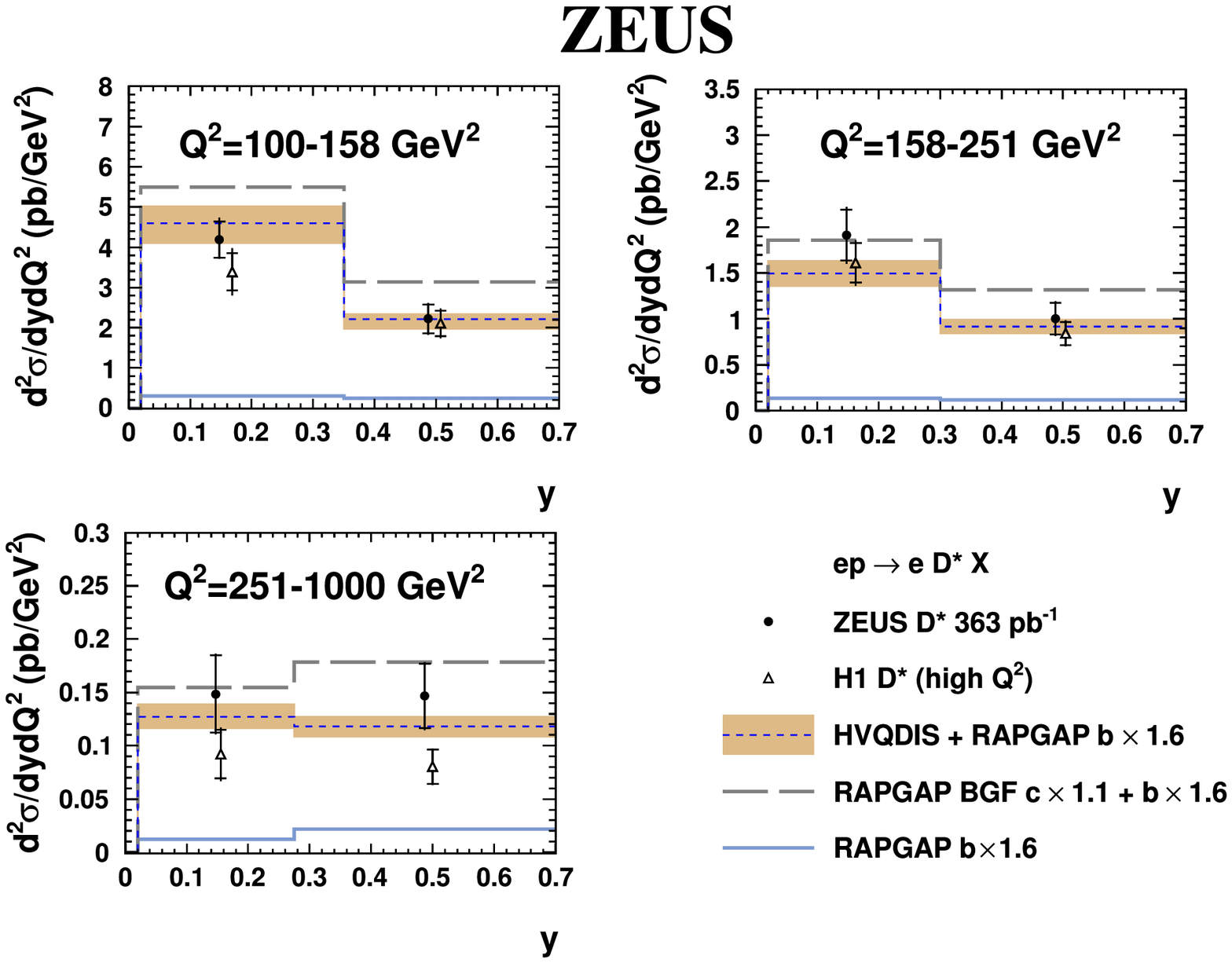}
    \end{center}
    \caption
    {
      \textit{Double-differential $D^{*\pm} $ cross sections as a function of $Q^{2}$ and $y$ 
        for $100<Q^{2}<1000$ GeV\,$^{\,2}$ (filled circles).  The  measurements from the H1 collaboration
        (empty triangles) are also shown~\protect\cite{Aaron:2009jy}. Other details as in Fig.~\protect\ref{f:single_2}.}
    }
    \label{f:double_2}
    \vfill
\end{figure}

%

\begin{figure}[htp]
\begin{center}
\includegraphics[width=0.9\textwidth]{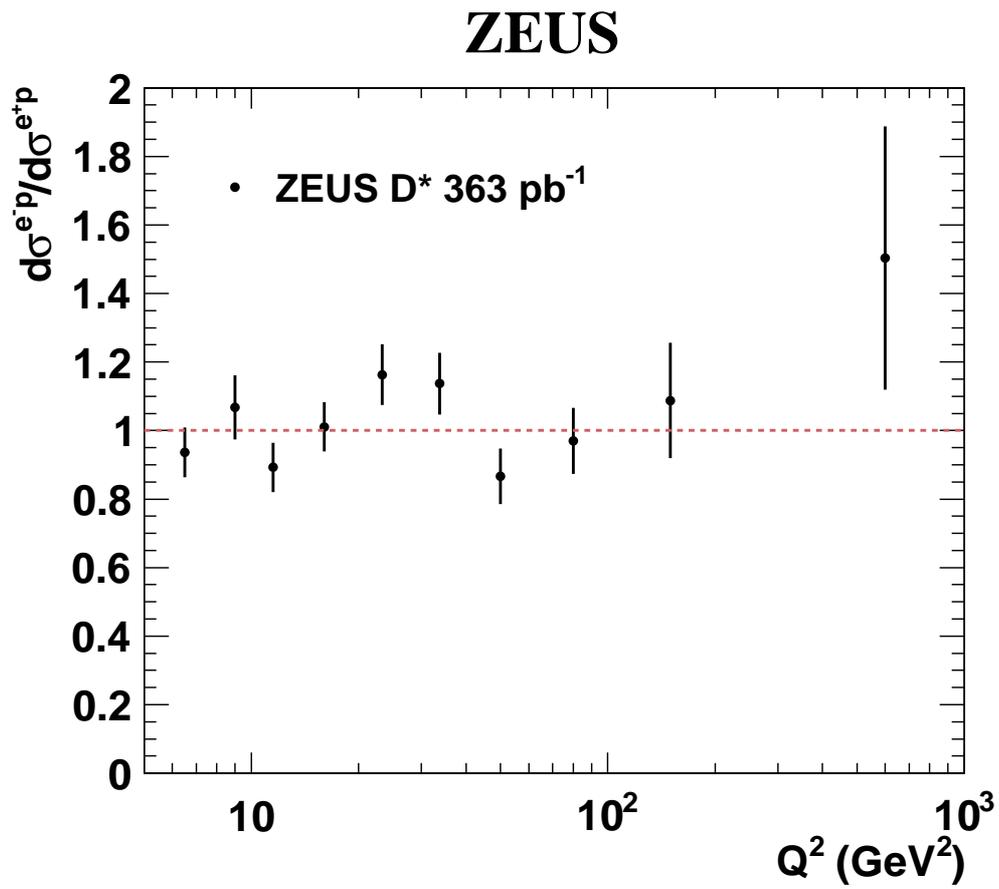}
\end{center}
\caption{
\textit{Ratio of $e^-p$ to $e^+p$ visible $D^{*\pm}$ cross
    sections as a function of $Q^2$.  Only statistical uncertainties are
    shown. Bin boundaries are as in Table~\ref{tab:xs_q2}.}
}
\label{f:epm}
\end{figure}

%


\begin{figure}[htp]
\begin{center}
\includegraphics[width=1.0\textwidth]{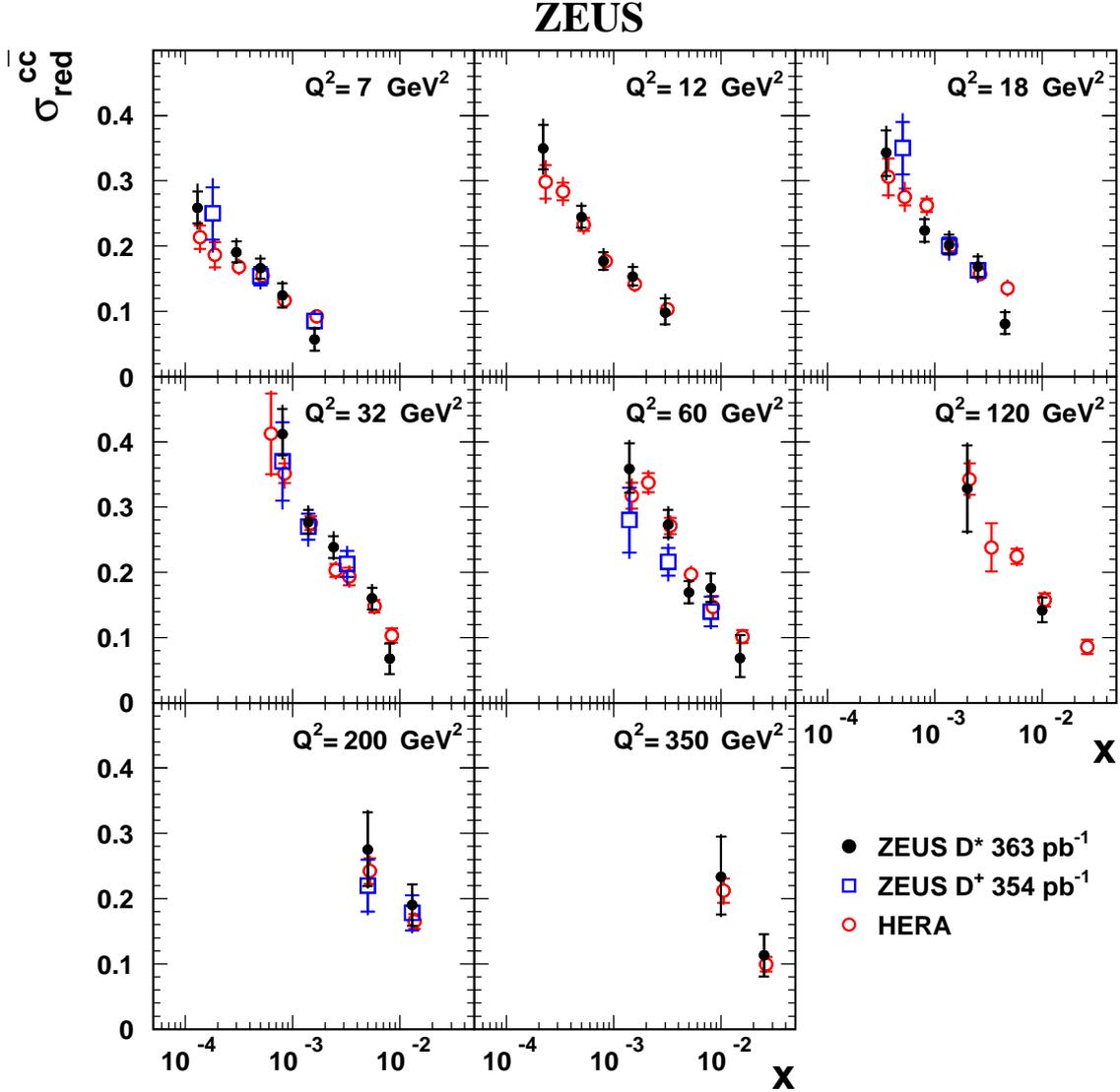}
\end{center}
\caption{ \textit{Reduced charm cross sections from $D^{*\pm}$ (filled circles) compared
to the ZEUS $D^+$ measurement~\protect\cite{zeus:dplus:2012} (empty squares) 
and the combination of previous HERA results~\protect\cite{h1zeus} 
(empty circles).  The outer error bars include
experimental and theoretical uncertainties added in quadrature. The inner error bars in the
ZEUS $D^*$ and $D^+$ measurements show the experimental uncertainties. The
inner error bars of the combined HERA data represent
the uncorrelated part of the uncertainty.}
}
\label{f:sigred_1}
\end{figure}

\begin{figure}[htp]
\begin{center}
\includegraphics[width=1.0\textwidth]{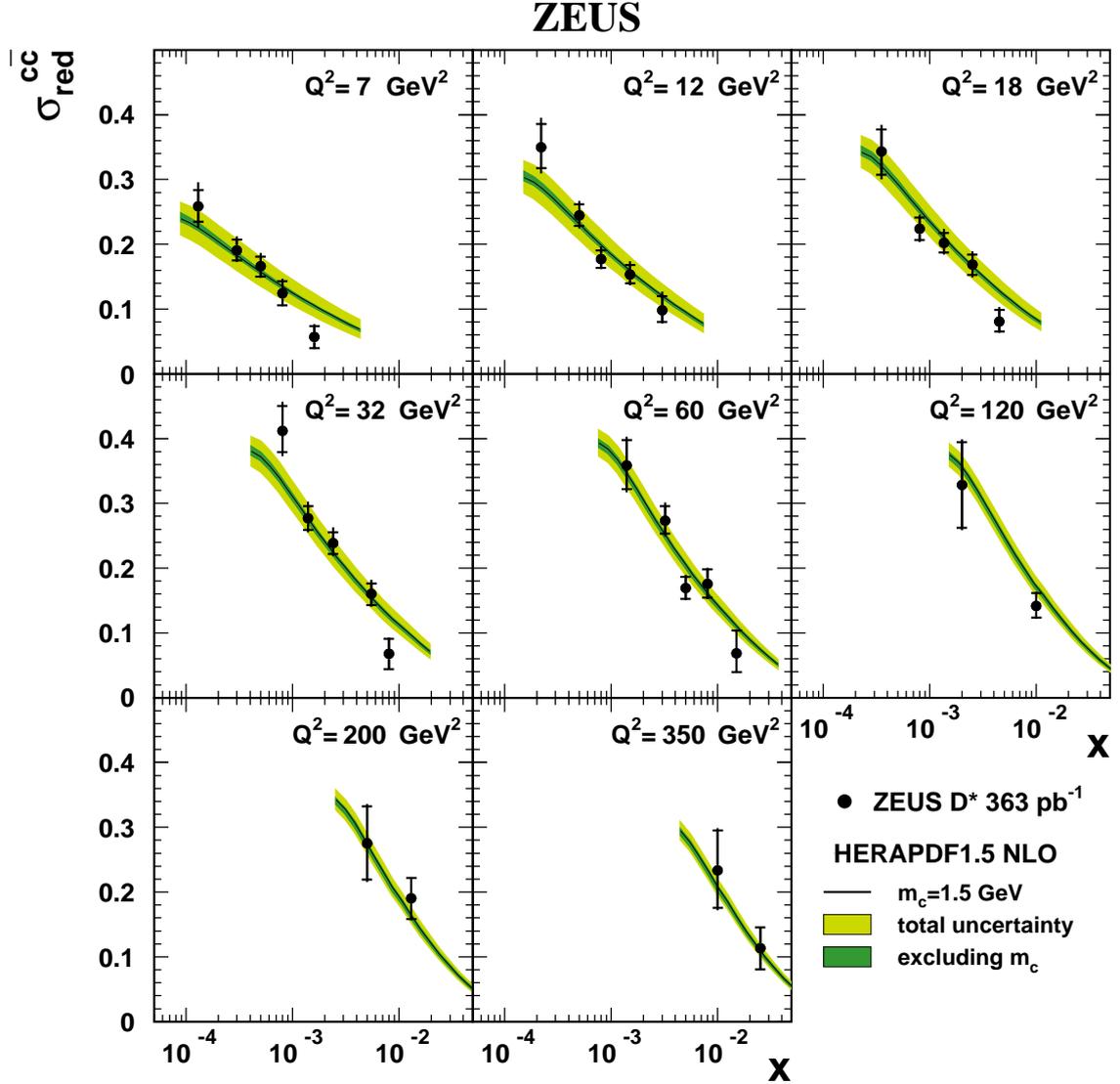}
\end{center}
\caption{
\textit{Reduced charm cross sections (filled circles) compared
to a GM-VFNS calculation based on HERAPDF1.5 parton
densities. The inner error bars show the experimental uncertainties and the outer error bars show the experimental and theoretical uncertainties added in quadrature. The outer bands on the
HERAPDF1.5 predicition show the total uncertainty while
the inner bands correspond to the sum in quadrature of all  
uncertainties excluding the charm-mass variation.}
}
\label{f:sigred_2}
\end{figure}